\documentclass[smallcondensed]{svjour3}

\usepackage{hyperref,xcolor,graphicx,amsmath,natbib}
\begin{document}

\title{Coronal heating by MHD waves}


\author{Tom Van Doorsselaere		\and
        Abhishek K. Srivastava 		\and
	Patrick Antolin			\and
	Norbert Magyar			\and
	Soheil Vasheghani Farahani	\and
	Hui Tian			\and
	Dmitrii Kolotkov		\and
	Leon Ofman			\and
	Mingzhe Guo			\and
	I\~nigo Arregui			\and
	Ineke De Moortel		\and
	David Pascoe			
	% anfinogentov needed?
}


\institute{Tom Van Doorsselaere \at Centre for mathematical Plasma Astrophysics, Department of Mathematics, KU~Leuven, Celestijnenlaan 200B bus 2400, B-3001 Leuven, Belgium\\ \email{tom.vandoorsselaere@kuleuven.be}
	\and
	Abhishek K. Srivastava \at Department of Physics, Indian Institute of Technology (BHU), Varanasi-221005, India
	\and
	Patrick Antolin	\at Department of Mathematics, Physics and Electrical Engineering, Northumbria University, Newcastle Upon Tyne, NE1 8ST, United Kingdom
	\and
	Norbert Magyar \at Centre for Fusion, Space and Astrophysics, Physics Department, University of Warwick, Coventry CV4 7AL, UK
	\and
	Soheil Vasheghani Farahani \at Department of Physics,Tafresh University, Tafresh 39518 79611, Iran
	\and
	Hui Tian \at School of Earth and Space Sciences, Peking University, Beijing 100871, China \\ Key Laboratory of Solar Activity, National Astronomical Observatories, Chinese Academy of Sciences, Beijing 100012, China
	\and
	Dmitrii Kolotkov \at Centre for Fusion, Space and Astrophysics, Physics Department, University of Warwick, Coventry CV4 7AL, UK\\
	Institute of Solar-Terrestrial Physics SB RAS, Irkutsk 664033, Russia
	\and
	Leon Ofman \at Department of Physics, Catholic University of America, Washington, DC, USA\\ NASA Goddard Space Flight Center, Greenbelt, MD, USA
	\and
	Mingzhe Guo \at Institute of Space Sciences, Shandong University, Weihai 264209, People's Republic of China \\ Centre for mathematical Plasma Astrophysics, Department of Mathematics, KU~Leuven, Celestijnenlaan 200B bus 2400, B-3001 Leuven, Belgium	
	\and
	I\~nigo Arregui	\at Instituto de Astrofísica de Canarias, 38205, La Laguna, Tenerife, Spain\\ Departamento de Astrofísica, Universidad de La Laguna, 38206, La Laguna, Tenerife, Spain
	\and
	Ineke De Moortel \at School of Mathematics and Statistics, University of St Andrews, North Haugh, St Andrews, KY16 9SS, UK\\ Rosseland Centre for Solar Physics, University of Oslo, PO Box 1029 Blindern, NO-0315 Oslo, Norway
	\and
	David Pascoe \at Centre for mathematical Plasma Astrophysics, Department of Mathematics, KU~Leuven, Celestijnenlaan 200B bus 2400, B-3001 Leuven, Belgium
}

\date{Received: date / Accepted: date}

\maketitle

\begin{abstract}
	The heating of the solar chromosphere and corona to the observed high temperatures, imply the presence of ongoing heating that balances the strong radiative and thermal conduction losses expected in the solar atmosphere. It has been theorized for decades that the required heating mechanisms of the chromospheric and coronal parts of the active regions, quiet-Sun, and coronal holes
are associated with the solar magnetic fields. However, the exact physical process that transport and dissipate the magnetic energy which ultimately leads to the solar plasma heating are not yet fully understood. The current understanding of coronal heating relies on two main mechanism: reconnection and MHD waves that may have various degrees of importance in different coronal regions. In this review we focus on recent advances in our understanding of MHD wave heating mechanisms. First, we focus on giving an overview of observational results, where we show that different wave modes have been discovered in the corona in the last decade, many of which are associated with a significant energy flux, either generated in situ or pumped from the lower solar atmosphere. Afterwards, we summarise the recent findings of numerical modelling of waves, motivated by the observational results. Despite the advances, only 3D MHD models with Alfv\'en wave heating in an unstructured corona can explain the observed coronal temperatures compatible with the quiet Sun, while 3D MHD wave heating models including cross-field density structuring are not yet able to account for the heating of coronal loops in active regions to their observed temperature.
\end{abstract}

%\linenumbers

\section{Introduction}
Coronal heating is a long-standing problem. It is quite clear that the energy for the hot corona comes from the convective motions of the solar photosphere, and that the magnetic field plays a key role in it. However, how the energy is transported and dissipated is still not fully understood. Proposed heating mechanisms are classified based on the comparison of the convective time scales and the Alfv\'en transit time in the corona. Slow driving of the magnetic field  that produces reconnection and energy release in current sheets and null points, is referred to as DC heating mechanisms \citep{1986GApFD..35..277P}. Fast driving of the magnetic field, is known as MHD wave heating \citep[e.g.][]{1947MNRAS.107..211A} and  AC heating mechanisms. \par
DC heating mechanisms consider the slow stressing of the coronal magnetic field. It leads to current sheets where Ohmic dissipation is at work, or to Parker's idea of nanoflares. In the latter idea, the magnetic field is tangled because of the photospheric motions, so that it evolves into a non-potential state. This magnetic energy is then released into the plasma by reconnection, resulting in localised heating and thus nanoflares. DC heating mechanisms have received a lot of attention over the years in numerical modelling and observations \citep[e.g.][]{gudiksen2005,reep2013,rempel2017,warnecke2017,Sri19}. \par
In this review, we focus on AC heating mechanisms. In these heating mechanisms, the convective motions launch disturbances that travel into the corona, usually in the form of magnetohydrodynamic (MHD) waves. In the corona, the wave energy has to be dissipated to heat the plasma. This is non-trivial, because many wave modes damp only on resistive time scales in a homogeneous plasma, and damping times are proportional to the magnetic Reynolds number. In the solar corona, the Reynolds number is very large, on the order of $10^{14}$, which would produce unrealistically long heating times, compared to the coronal cooling  timescale. Cross-field inhomogeneity enables physical processes that produce a cascade of wave energy to small spatial scales where dissipative processes may act and heat the plasma more rapidly. In recent years, there has been a drive towards (1) observational characterisation of wave energy content in the corona, and (2) numerical modelling of wave heating, focusing on the energy input, the energy propagation and the energy dissipation. Here, we aim to give an overview of these recent results. \par
For material beyond the current review, we refer the reader to, for example,  \citet{aschwanden2019}.

\subsection{Brief historical overview}
The idea for heating the solar corona by MHD waves has been around for more than half a century. An overview of the early ideas before the 80s can be found in \citet{kuperus1981}. In those years, it was realised that it is necessary to generate small scales in order to damp the wave energy in a timely manner. \citet{heyvaerts1983} developed the theory of phase mixing, while resonant absorption \citep{chen1974} was applied to the corona for the first time by \citet{ionson1978}. Subsequently, the MHD wave energy input and transmission into coronal loops was studied in key papers such as \citet{hollweg1984}. Later on, the efficiency of the heating by resonant absorption was calculated in 1D numerical models of loops \citep[][among others]{poedts1990b}, and in 3D models \citep[][and follow-up works]{ofman1994}, even resulting in forward modelled coronal loops \citep{belien1996}, motivated by early high resolution soft X-ray observations of coronal loops that became available from the Yohkoh satellite. \par
In hindsight, it is amazing that these early papers worked so well and were so relevant, given that no direct observational evidence was available back then on the presence of MHD waves in the solar corona. Indeed, although substantial, only indirect evidence existed of the potentially important role of MHD waves in the solar atmosphere. Such evidence was based on the strong emission and broad non-thermal line widths in the upper chromosphere, transition region and corona with observations from Skylab \citep{Feldman_1988JOSAB...5.2237F} and HRTS \citep[High-Resolution Telescope and Spectrograph,][]{Dere_1993SoPh..144..217D}. This changed dramatically with the launch of SOHO (Solar and Heliospheric Observatory) and TRACE (Transition Region And Coronal Explorer). Data from the former were used to show the presence of slow waves in the corona \citep{Chae_etal_1998ApJ...505..957C,1999ApJ...514..441O,berghmans1999}, while data from the latter revealed the presence of transverse, post-flare loop oscillations \citep{nakariakov1999,schrijver1999,aschwanden1999}. An overview of the contextualisation of earlier models with those observational findings is given in \citet{walsh2003,2005SSRv..120...67O}. \par
LCR models (named after the usual symbols in electric components for inductance L, capacitance C and resistance R) were also developed extensively for MHD waves in coronal loops \citep[see][and reference therein]{stepanov2012}. In such models, the magnetic twist in the loop behaves as a current system with its source in the photosphere and equivalent electric resistivity and inductance in the corona \citep[see e.g.,][]{spicer1977,carlqvist1979}. However, it is still unclear how this LCR model relates to MHD models of coronal loop oscillations \citep[based on e.g.][]{edwin1983}. Moreover, since no development of these models was made in the last years, we will omit them from this review.

\subsection{Observational motivation}
% * <Tom Van Doorsselaere> 17:23:57 18 Feb 2020 UTC+0100:
% Patrick, can you make a first attempt at this section?
Starting with the advent of high resolution space-based observations of the solar corona in soft X-ray and EUV  in the 90s, there is now an avalanche of new MHD wave modes detected in the solar corona, as described in detail by \citet{demoortel2012b}. Their consequences for heating the corona are discussed in \citet{arregui2015}. 

Another major paradigm change in our perception of MHD waves in the solar atmosphere came from ground-based observations with \textit{CoMP} \citep{tomczyk2007}, which showed the omnipresence of these waves in the solar corona, and spectroscopic and imaging from space with \textit{Hinode}, which, thanks to its high resolution, allowed to better quantify the amount of energy available for the corona \citep{DePontieu_2007Sci...318.1574D}. Moreover, \textit{SDO/AIA} revealed the existence of many new wave modes and made more detailed observations of previously observed coronal waves \citep[for a review, see][]{LO14}.

The improvement in spatial, spectral and temporal resolution of instrumentation also meant that MHD waves could be characterised in terms of their slow, fast or Alfv\'en nature, as well as their wave numbers \citep[see the review by][]{demoortel2012b}. Indeed, high spatial resolution and high sensitivity allows to distinguish transverse motions of the waveguides as well as the compressibility of the gas from density diagnostics, while high spectral and temporal resolution allows to relate these quantities to the evolution of the Doppler and non-thermal velocities, thereby determining the 3D motion of the plasma produced by the waves \citep[see e.g. ][chapter 6.1]{Fujimura_Tsuneta_2009ApJ...702.1443F,Kitagawa_Yokoyama_2010ApJ...721..744K,2016GMS...216..395W,Hinode_10.1093/pasj/psz084}.

Lastly, multi-wavelength observations with instruments such as those of \textit{Hinode} and \textit{IRIS}, or through coordinated observations with space and ground-based observatories allowed to simultaneously scan several layers of the solar atmosphere. Although a complicated task, this provides the propagation history of the wave, which is essential to determine wave processes such as reflection and refraction, and in particular mode conversion and dissipation of MHD waves for coronal heating \citep{arregui2015}.

\section{Observations}
Below, we describe recent detections of wave power in the solar atmosphere. We summarise the energy fluxes in Table~\ref{tab:flux} to give a clear overview.
\begin{table}
	\caption{Overview of detected wave energy fluxes. The first column shows the structure in the solar atmosphere with the observed wave mode (second column) and the observing instrument (third column). The fourth column shows the estimated energy flux, as found by the reference mentioned in the fifth column. If a cell is left empty, the value from the previous line are meant.}
	\label{tab:flux}
	\begin{tabular}{lllcl}
		\hline\noalign{\smallskip}
		Structure & Wave mode & Instrument & Energy flux (W m$^{-2}$) & Reference  \\
		\noalign{\smallskip}\hline \hline \noalign{\smallskip}
		coronal arcade & propagating kink & CoMP & 100 & \citet{tomczyk2007} \\ \hline
		coronal funnels & quasi-periodic fast wave trains & AIA & $(0.1-2.6)\times 10^4$ & \citet{Liu11}\\
		&& simulations & $3.7\times 10^5$ & \citet{Ofm11} \\
		&& AIA & $1.8\times 10^2$ & \citet{OL18} \\ \hline
		magnetic pores & Alfv\'en waves & ROSA & 1.5$\times$10$^{4}$ (locally) & \citet{2009Sci...323.1582J}\\
		& & & 240 (globally averaged) & \\ 
		& propagating slow sausage & DST & 3.5$\times$10$^{4}$ & \citet{2015ApJ...806..132G}\\ \hline
		fibrils & propagating kink & ROSA & $4.3\times 10^3$ & \citet{2012NatCo...3.1315M}\\
		& propagating fast sausage & & $1.17\times 10^4$\\ \hline
		spicules & Alfv\'en waves & SST & $10^5$ & \citet{2017NatSR...743147S}\\
		& kink & & $10^2-10^4$ & this review, based on \citet{DePontieu_2012ApJ...752L..12D}\\
		\noalign{\smallskip}\hline
	\end{tabular}
\end{table}

\subsection{Impulsively excited standing kink waves}\label{sec:standing}
Often flares and low coronal eruptions (LCEs) are seen to excite standing kink waves in coronal loops \citep{2015A&A...577A...4Z}. \citet{2016A&A...585A.137G} and \citet{2019ApJS..241...31N} performed a statistical study of these types of events and found that the loops oscillate with an amplitude between 1-10\ Mm and periods between 1-28\ min. While the coronal impulsive event that causes the oscillations is apparent, the exact mechanism for their excitation is not well understood. Several possibilities were investigated in the literature: internal and external (gas or magnetic) pressure drivers \citep{terradas2007,2008ApJ...682.1338M, 2009A&A...505..319P,2009AnGeo..27.3899S,2014ApJ...784..101P}, loop contraction \citep{2015A&A...581A...8R,2017A&A...607A...8P}, or collision of flows \citep[][even though these would result in propagating waves]{Antolin_2018ApJ...861L..15A,Pagano_2019AA...626A..53P}. More information about observations of impulsively excited standing waves can be found in the review by \citet{nakariakov2020}.

\begin{figure}
	\includegraphics[width=\linewidth]{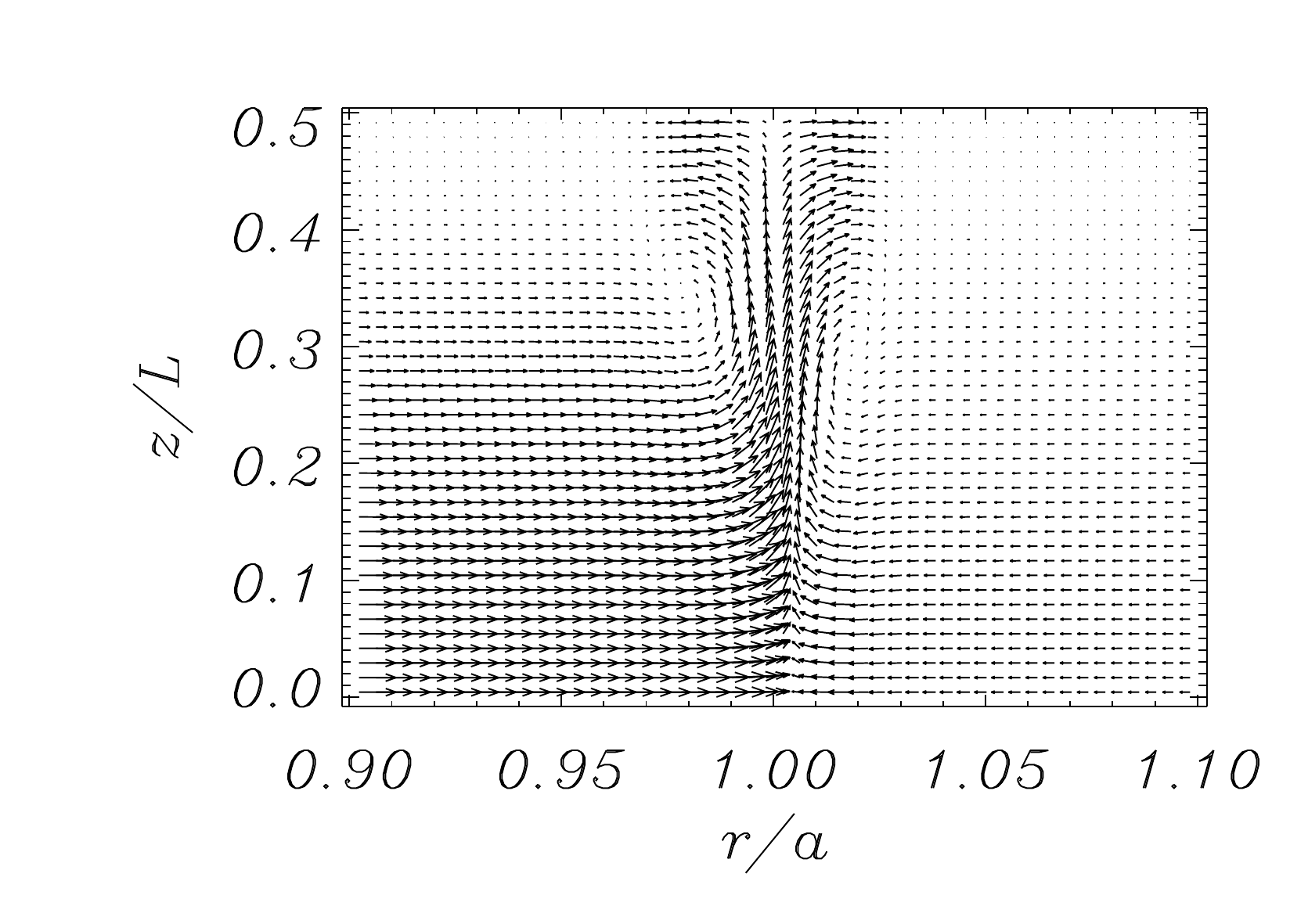}
	\caption{Spatial distribution of the Poynting vector in the ($r, z$)-plane around the resonant position at $r/a=1$ for a standing kink mode in a loop with a density contrast of 10 and a non-uniform layer of length $l/a=0.2$. The magnetic Reynolds number is $R_{\rm m}=10^6$. Figure modified from \citet{arregui11}.}
	\label{fig:poynting}
\end{figure}

The energy available in impulsively excited standing waves and the fraction that is dissipated remains to be characterised from observations. An energy analysis based on bulk plane Alfv\'en waves is too simplistic because kink mode energy is localised in space \citep{goossens2013}. Magnetic and plasma structures act as frequency filters, trapping part of the available energy and distributing it along and across the coronal field.  Wave energy propagation, once filtered by the structure is localised in space. Energy flows into the resonance because of the jump in the radial component of the Poynting vector \citep{arregui11}, as shown in Fig.~\ref{fig:poynting}. Then, it propagates and dissipates along the field in a way determined by the density profile and dissipative coefficients.

\cite{terradas2018b} considered each part of the sequence of physical processes that would enable to heat a typical coronal waveguide from the energy contained in a typical transverse oscillation. Each part of the sequence has its characteristic time and spatial scales. The damping time is determined by the cross-field plasma and field structuring. The energy cascade to small scales is determined by a phase mixing length that also depends on the cross-field variation of the Alfv\'en speed. The onset of resistive dissipation and its duration depend on the Reynolds number and the cross-field plasma variation.  Considering typical values of these quantities we cannot expect resistive diffusion to operate during the oscillation process. Any observational evidence about wave heating by resistive damping of impulsively excited standing waves will come from the observation of indirect consequences.  According to \citep{terradas2018b}, for a loop displacement of the order of the
radius and typical loop
parameters, all the kinetic and magnetic energy
 of a typically observed single kink mode is of the
 order of $10^{19}$ J. Simple energy conservation calculations indicate that, even if we were able to concentrate in a typical resonant layer and transform into internal energy all the kinetic and magnetic energy of a typically observed single kink mode, we could just obtain a temperature increase of about 10$^5$ K \citep{terradas2018b}. 

When an ensemble of mini-tubes are considered, filling factors can be employed \citep{vd2014}, which leads to an energy flux reduction directly related to the filling factor. Although resonant damping and mode coupling are robust in coronal loop models with rather arbitrary continuous plasma distributions, the multi-strand structure of loops could be quickly destroyed due to instabilities in individual strands and their interaction \citep{magyar2016}.

\subsection{CoMP waves}

\begin{figure*} 
\centering {\includegraphics[width=0.7\textwidth, angle=-90]{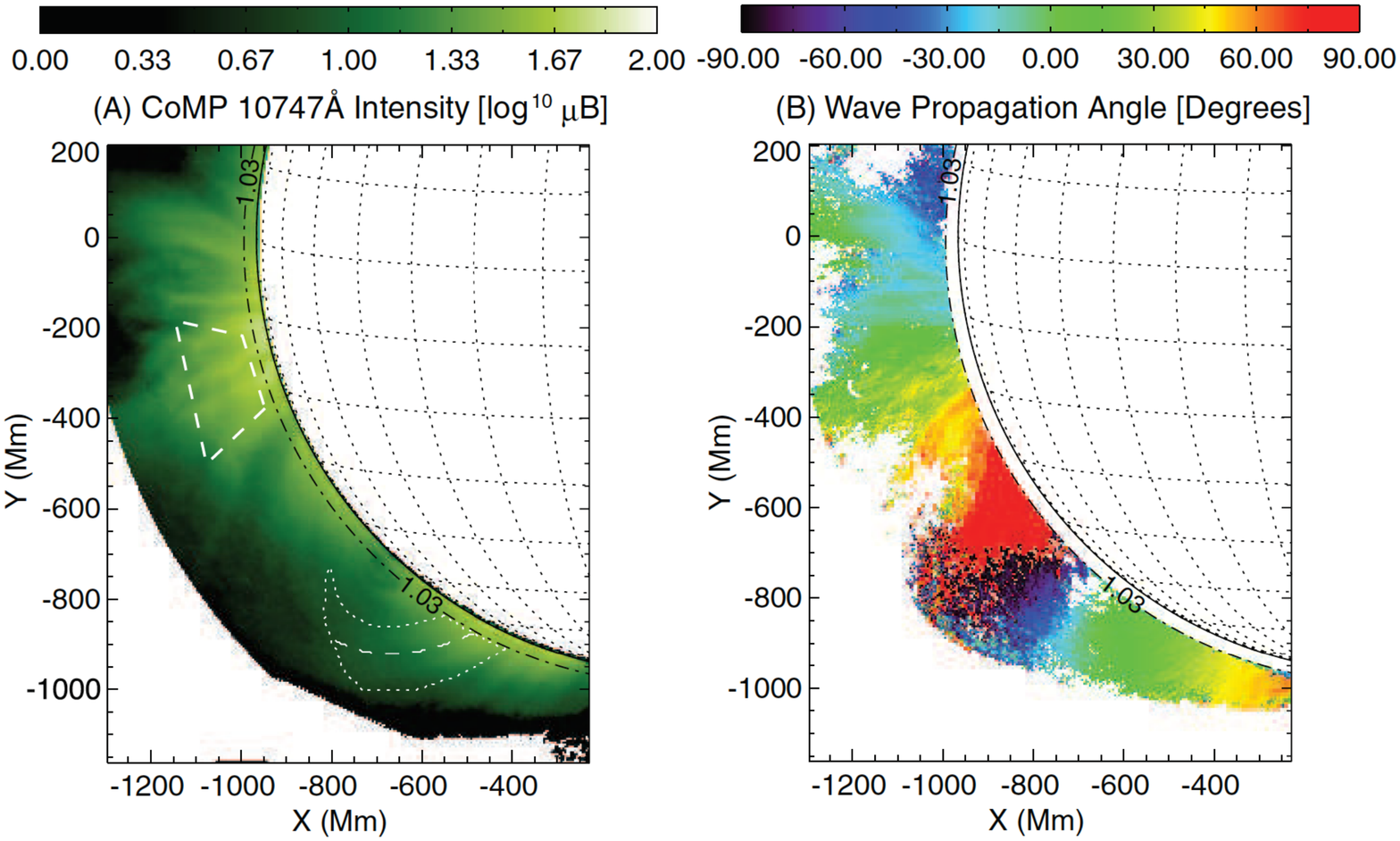}} 
\caption{ Images of the Fe~{\sc{xiii}} 10747 {\AA}~line intensity and the wave propagation angles derived from the Doppler shift of Fe~{\sc{xiii}} 10747 {\AA} in part of the corona observed by CoMP. Adapted from \cite{tomczyk2009}.} \label{f1}
\end{figure*}

Prevalent propagating waves were reported \citep{tomczyk2007,2011Natur.475..477M} from observations  of both the Coronal
Multi-channel Polarimeter \citep[CoMP,][]{tomczyk2008} and the Atmospheric Imaging Assembly \citep[AIA,][]{lemen2012} on board the Solar Dynamics Observatory (SDO). Using CoMP observations, \citet{tomczyk2007} found ubiquitous upward propagating disturbances along off-limb coronal loops in the Doppler shift of the Fe~{\sc{xiii}} 10747 {\AA}~line (Figure~\ref{f1}). Since the Doppler shift refers to the motion along the line of sight (LOS) and the magnetic field lines in the off-limb corona are largely perpendicular to the LOS, these disturbances are essentially signatures of transverse MHD waves. The propagation speed was found to be from a few hundred km~s$^{-1}$ to 2000 km~s$^{-1}$, which is on the order of the coronal Alfv\'en speed. Based on these characteristics, \citet{tomczyk2007} and \citet{tomczyk2009} interpreted these transverse waves as Alfv\'en waves. They found a power-law spectrum of the Doppler shift with a spectral index of about 1.5, consistent with isotropic MHD turbulence. The power spectrum of the observations shows a peak around 3.2 mHz (period of 5 minutes). The velocity amplitude is only 0.5 km~s$^{-1}$, and the energy flux of these waves appears to be at least three orders of magnitude lower than that required to balance the radiative losses of the quiet corona. The low-amplitude is likely due to the low spatial resolution of CoMP ($\sim$9$^{\prime\prime}$) and the line-of-sight integration resulting in spatial averaging of the waves. Similar results were also found in open-field regions \citep{Morton2015}. By carefully analyzing the displacement of coronal structures observed by the much higher-resolution AIA instrument ($\sim$1.5$^{\prime\prime}$), \citet{2011Natur.475..477M} found much larger velocity amplitudes, i.e., $\sim$20 km~s$^{-1}$, for these transverse waves. They estimated the energy flux of the waves as $\sim 100\ {\rm W\,m}^{-2}$ and found that the waves have sufficient energy to power the quiet corona and fast solar wind. However, the observed wave energy flux is still much lower than that required to heat the active region corona ($\sim$2000~W\,m$^{-2}$).

The interpretation of these waves as Alfv\'en waves was controversial. As argued by \citet{vd2008}, in cylindrical plasma structures these waves appear to be more appropriately interpreted as fast-mode kink waves rather than Alfv\'en waves. However, others claimed that these waves could be called Alfv\'enic waves as the major restoring force for these waves is definitely magnetic tension \citep{Goossens2009,2011Natur.475..477M}, they carry parallel vorticity \citep{goossens2012} and they are inherently coupled to the local azimuthal Alfv\'en waves \citep{pascoe2010}. This mode is also called surface Alfv\'en wave or resonantly damped kink mode. At the end, the heating will come from azimuthal Alfv\'en wave dissipation. The global eigen-mode just plays the role of trapping/chanelling the energy from around/below. A well defined structure is not needed, as shown by \citet{terradas2008,2010ApJ...713..651R,pascoe2011}. \\
Despite the debate on the nature and terminology for these waves, the discovery of these propagating transverse waves has caught the attention of many researchers in the field. There are at least two reasons for this. First, they allow us to map the global coronal magnetic field based on the technique of coronal seismology \citep{magyar2018,2019ApJ...884L..40A,Yang2020a,Yang2020b}. Second, their ubiquity means they could potentially play an important role in coronal heating. The observed waves show a discrepancy between the outward and inward wave power \citep{tomczyk2009}, a property that resonantly damped wave models seem to capture well \citep{2010ApJ...718L.102V, pascoe2010}, and that is also observed in-situ in the solar wind \citep[see the review by][]{2013LRSP...10....2B}. A recent study by \citet{2020A&A...640L..17M} considers a distinct power generated at loop foot-points as an additional source of discrepancy. It shows that, if present, it would affect obtaining quantitative evidence for resonant damping. 

The origin of the waves detected by CoMP has recently been linked to the Sun's internal acoustic oscillations, and in particular to p-modes. Indeed, \citet{Morton_2019NatAs.tmp..196M} have shown that there is a distinct excess of wave power at the usual p-mode frequency of $3-5$~mHz, across the solar cycle and different coronal structures (or regions) in the Sun \citep{Morton_2016ApJ...828...89M}. The authors further show that the usual Alfv\'enic fluctuations observed with SDO/AIA \citep{2011Natur.475..477M} have the same characteristics in their power spectra as the CoMP waves. Evidently, p-modes contain orders of magnitude more power than required for coronal heating. The physical mechanism proposed to explain the ubiquity of these waves is the double mode conversion process of p-modes during their propagation across the low atmospheric layers, during which their energy is transferred to Alfv\'enic modes \citep{Cally_2011ApJ...738..119C,Felipe_2012ApJ...758...96F,2017MNRAS.466..413C}. A strong consequence of this is that solar and stellar coronae have a non-negligible energy source in their internal acoustic oscillations. Although direct observations of this process are still lacking, these results also suggest that p-modes may play a role in the onset of standing or propagating Alfv\'enic modes.

\subsection{Decayless oscillations}

\begin{figure*} 
\centering {\includegraphics[width=0.7\textwidth]{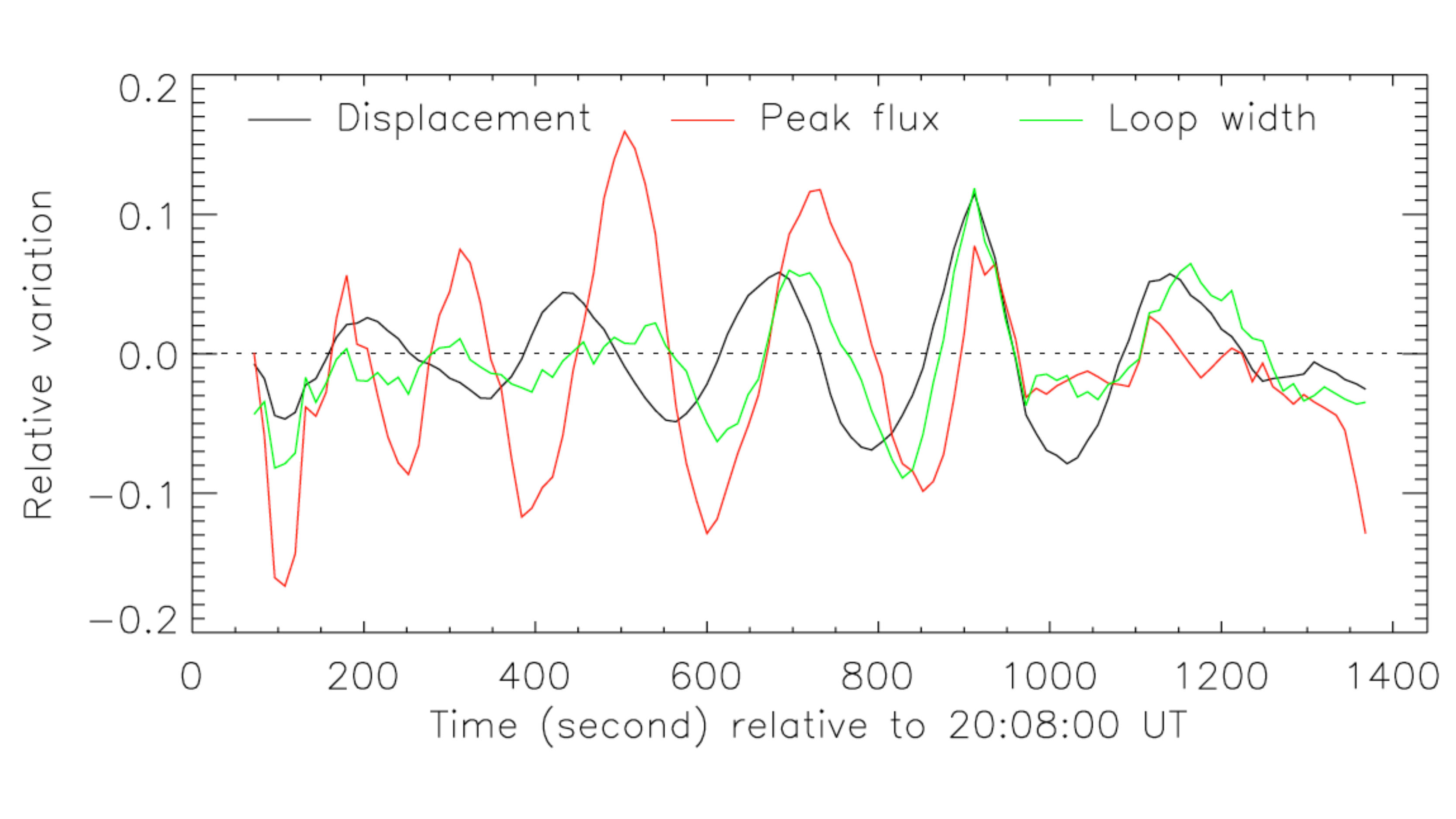}} 
\caption{ Decayless oscillations of coronal loops seen in the loop displacement, peak flux and loop width estimated from the AIA 171 {\AA}~images. Adapted from \cite{Wang2012}.} \label{f2}
\end{figure*}

\begin{figure*} 
\centering {\includegraphics[width=0.6\textwidth, angle=-90]{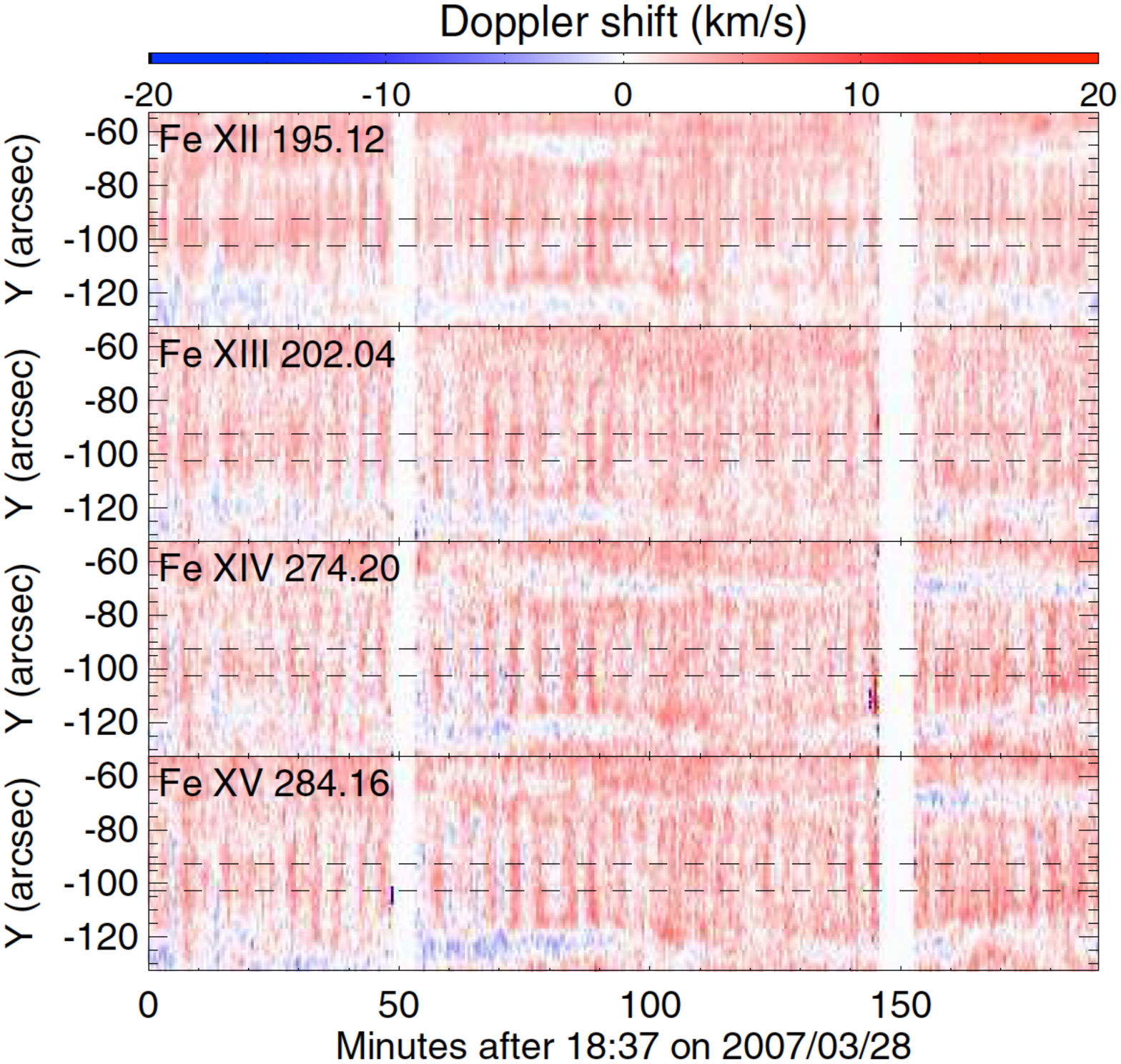}} 
\caption{ Decayless oscillations of quiet coronal loops seen in the Doppler shift of four coronal emission lines. The Y-axis is roughly aligned with the corona loop. Adapted from \cite{Tian2012}.} \label{f3}
\end{figure*}

Decayless oscillations normally refer to standing waves without obvious observed damping in coronal loops. In coronal lines\footnote{Small amplitude decayless oscillations were also observed in a prominence with \textit{Hinode}/SOT \citep{Ning_2009AA...499..595N}, and of significant and increasing amplitude in a coronal loop with rain \citep{Antolin_Verwichte_2011ApJ...736..121A}. Due to the presence of chromospheric material and the associated processes specific to the formation of prominences and coronal rain, the processes responsible for such oscillations may likely be different than those observed in coronal lines.}, these oscillations were first reported independently by \citet{Wang2012} and \citet{Tian2012} through imaging and spectroscopic observations, respectively. The transverse oscillations reported by \citet{Wang2012} from SDO/AIA observations appear to be triggered by a coronal mass ejection. They lasted for more than ten cycles and even revealed growing amplitudes (Figure~\ref{f2}). \cite{Tian2012} performed a survey of decayless (referred to as ``persistent'' in their paper) Doppler shift oscillations using three-month spectroscopic observations from the EUV Imaging Spectrometer \citep[EIS,][]{Culhane2007} on board Hinode. They found that such decayless oscillations, with a period of 3--6 minutes and a velocity amplitude of 1--2 km~s$^{-1}$, are very common in quiet coronal loops and can be observed in several coronal emission lines formed in the temperature range of 1.3--2 MK (Figure~\ref{f3}). These oscillations generally reveal no obvious damping during the whole observed time interval, which often last for a few hours. They were interpreted as standing kink oscillations  by these authors. Since 2012, these decayless oscillations have been frequently reported for a variety of different observations \citep[e.g.,][]{Nistico2013,Nistico2014}, and it was found that their periods scale with the loop length \citep{Anfinogentov2015}, strongly supporting the standing wave interpretation.

The decayless behavior of these waves suggests a continuous supply of energy to the system. Several different ideas have been proposed to maintain these decayless or persistent oscillations. For instance, \citet{Nakariakov2016} interpreted these decayless kink oscillations as a self-oscillatory process \citep[a process in which the driver is a consequence of the oscillation itself, see][]{jenkins2013}, driven by interaction between the loops and quasi-steady flows at the loop footpoints. \citet{Antolin_2016ApJ...830L..22A} have shown that the decayless pattern could also be an apparent effect produced by the combination of low instrumental resolution and the periodic brightening generated by TWIKH rolls (Transverse Wave Induced Kelvin-Helmholtz rolls, cf. Sec.~\ref{sec:kinkkhi}). In this interpretation, the decayless character reflects the low damping character of the azimuthal Alfv\'en waves resonantly coupled to the kink mode at the boundary layer. Another theory is through continued footpoint driving \citep{karampelas2017} providing an upward Poynting flux balancing the resonant and non-linear damping of the kink waves. Recent numerical experiments \citep{karampelas2019b,guo2019} reproduce the observations of decayless oscillations very well.  More recently, \citet{Afanasyev2020} developed a one-dimensional and time-dependent analytical model by considering kink oscillations of coronal loops driven by random motions of loop footpoints. They have managed to reproduce a number of observational facts about these decayless kink oscillations, such as the quasi-monochromaticity, period-length relationship and excitation of multiple harmonics.\par

\subsection{Available energy in quasi-periodic fast mode wave trains}
The high cadence and spatial resolution of SDO/AIA in EUV enabled the discovery of quasiperiodic fast mode propagating wave trains (QFPs) (see, \citet{Liu10,Liu11}, the review \citet{LO14}. The waves were observed to be associated with flares and propagate at a high speed of 1000-2000 km s$^{-1}$ with periods in the range of minutes and intensity variations $\delta I/I\sim5$\%. The waves were modeled first by \citet{Ofm11} using a 3D MHD model of a realistic bipolar active region structure and identified as fast-mode magnetoacoustic waves. Since the initial discovery, these waves were observed, studied, and modeled in many events associated with flares \citep[see, e.g.][]{Liu12,She12,She13,she17,She18,Yua13, Kum13, Nistico2014,Zha15,God16,Qu17,OL18}. Evidently, the QFPs carry some of the energy flux produced by the flares, and the observed dissipation of the QFP waves should result in coronal heating. The energy flux of the waves was estimated first using the parameters observed by SDO/AIA on 2010/08/01 by \citet{Liu11} and was found to be in the range $(0.1-2.6)\times 10^4$~W\,m$^{-2}$, which is of the same order as the steady-state heating requirement of active region loops.  The QFP wave energy flux estimate was based on the WKB approximation given by $E =\rho\delta v^2 v_{ph}/2\geq\rho (\delta I/I)^2 v_{ph}^3/8$ \citep{Asc04}, where the intensity of the line emission, $I\propto \rho^2$, and the phase speed $v_{ph}$ were determined directly from SDO/AIA observations  ($v_{ph} =1600$ km s$^{-1}$), and the number density is taken to be the typical coronal electron density $\sim 10^8$ cm$^{-3}$. The QFP wave trains lasted on the order of $\sim$ 0.5 hr and were repeatedly produced in the 2010/08/01  event. However, the divergence of the magnetic funnel in the active region  and apparent dissipation of the waves reduced the apparent energy flux away from the flaring source. 

 \citet{Ofm11} used the 3D MHD model results in an idealized bipolar active region of QFP wave trains combined with the WKB approximation to estimate the energy flux as $E=\rho \delta v^2 V_f /2 = 3.7\times 10^5$~W\,m$^{-2}$, where $\rho$ is the coronal density, $\delta v$ is the wave velocity amplitude, and $V_f$ is the fast magnetoacoustic speed. The model energy flux was an order of magnitude greater than the direct estimate by  \citet{Liu11}. However, the model estimate was based on idealized AR parameters in qualitative agreement with observations. In a recent study by \citet{OL18} of the double QFP event observed by SDO/AIA on 2013-5-22 the lower limit of the wave energy flux was estimated as $1.8\times 10^2$~W\,m$^{-2}$ high in the corona. The authors concluded that, taking into account coronal loop expansion, the wave energy flux could be at least one to two orders of magnitude higher near the source of the QFPs. 
 
 Statistical study of QFP waves indicates that these waves are quite common and often observed to be associated with C-class flares, although they can also result from stronger and weaker flares  \citep{Liu16}. Thus, combined with the challenging detection of these waves for lower energy flares, the contribution  of QFP wave trains to coronal heating and to the energy flux requirement may be more significant than initially estimated. Thus, the problem of QFP wave train heating parallels the coronal heating problem by flares, where small energy flares (i.e., undetected nanoflares) are required for coronal heating and possibly their associated QFP waves are not directly detected \citep{Asc04}.

\subsection{Observations and energy estimates of MHD waves in the lower solar atmosphere}

%------ Section about Energy Flux in the Lower Solar Atmosphere ------------------------------------------------------------------------------
%energy flux in lower atmosphere \citep{morton2012,moreels2015,grant2015}\\
%including work by Abhishek (high frequency waves) {\color{green} Abhishek}\\
The lower solar atmosphere (photosphere and chromosphere) plays an important role in the transmission of wave energy from the solar interior to the corona. Therefore, the assessment of wave energy content in this layer is crucial in understanding the coronal heating problem. Moreover, it is well established that the solar chromosphere requires substantially more energy flux than the corona (10$^{3}$--10$^{4}$ W\,m$^{-2}$) to compensate for its huge radiative losses \citep{1977ARA&A..15..363W}. 

Recent observational findings from high-resolution ground (e.g., 1m-Swedish Solar Telescope, ROSA -- Rapid Oscillations in the Solar Atmosphere) and space-based (e.g., IRIS -- Interface Region Imaging Spectrometer, SDO -- Solar Dynamics Observatory) observations reveal the presence of a high amount of energy flux associated with different MHD modes in a variety of magnetic structures coupling the various layers of the solar atmosphere. \par

In the photosphere, wave behaviour is found in pores and sunspots. \citet{2009Sci...323.1582J} reported the detection of Alfv\'en waves with periods of the order of 126-400 s associated with a large bright-point group. The energy flux associated with these wave modes is found to be 1.5$\times$10$^{4}$ W m$^{-2}$ in the chromosphere, which partially fulfills the chromospheric energy requirements. Using that 1.6\% of the solar surface is observed to be covered by bright-points and assuming that they support such torsional waves with a 42\% transmission coefficient (which may be on the high side), \citet{2009Sci...323.1582J} have estimated the global average energy in the corona as 240 W m$^{-2}$, which is sufficient to heat it locally. \\ 
\citet{2015ApJ...806..132G} have detected upwardly propagating slow sausage waves in magnetic pores, which initially carry an energy flux of 3.5$\times$10$^{4}$ W m$^{-2}$. The sausage wave energetics show a substantial decrease up to the chromosphere. These observations make it evident that magnetic pores transport waves to the higher layers, while also releasing energy in the local chromospheric plasma. However, it is not well quantified how much wave energy undergoes mode conversion, reflection or refraction. Aside from the frequently observed sausage modes, higher order magnetoacoustic oscillations ($m\geq 1$)  have also been observed in photospheric waveguides \citep[e.g.][and references cited there]{2017ApJ...842...59J, 2018ApJ...869..110S}, including body and surface modes \citep{2018ApJ...857...28K}. Furthermore, \citet{2018NatPh..14..480G} presented evidence of Alfv\'en wave heating of the chromospheric plasma in an active region sunspot umbra. They showed the presence of mode conversion and the formation of magnetoacoustic shocks.\\
In conclusion, the random buffeting motions in the photosphere generate many MHD modes in the photospheric magnetic flux tubes along with a sufficient amount of energy flux. However, one has to bear in mind that these magnetic flux concentrations constitute only a small part of the photosphere, and it is unclear if these wave motions also propagate to the corona in quiet regions, where such photospheric magnetic flux tubes are less prevalent.\par

%%%%%%%%%%%%%%%%%%%%%%%%%%%%%%%%%%%%%%%%%%%%%%%%%%%%%%%%%%%% Figure 5%%%%%%%%%%%%%%%%%%%%%%%%%%%%%%%%%%%%%%%%%%%%%%%%%%%%%%%%%%%%%%%%%%%% 
\begin{figure*} 
\hspace{-4.0cm}
\includegraphics[width=1.6\textwidth]{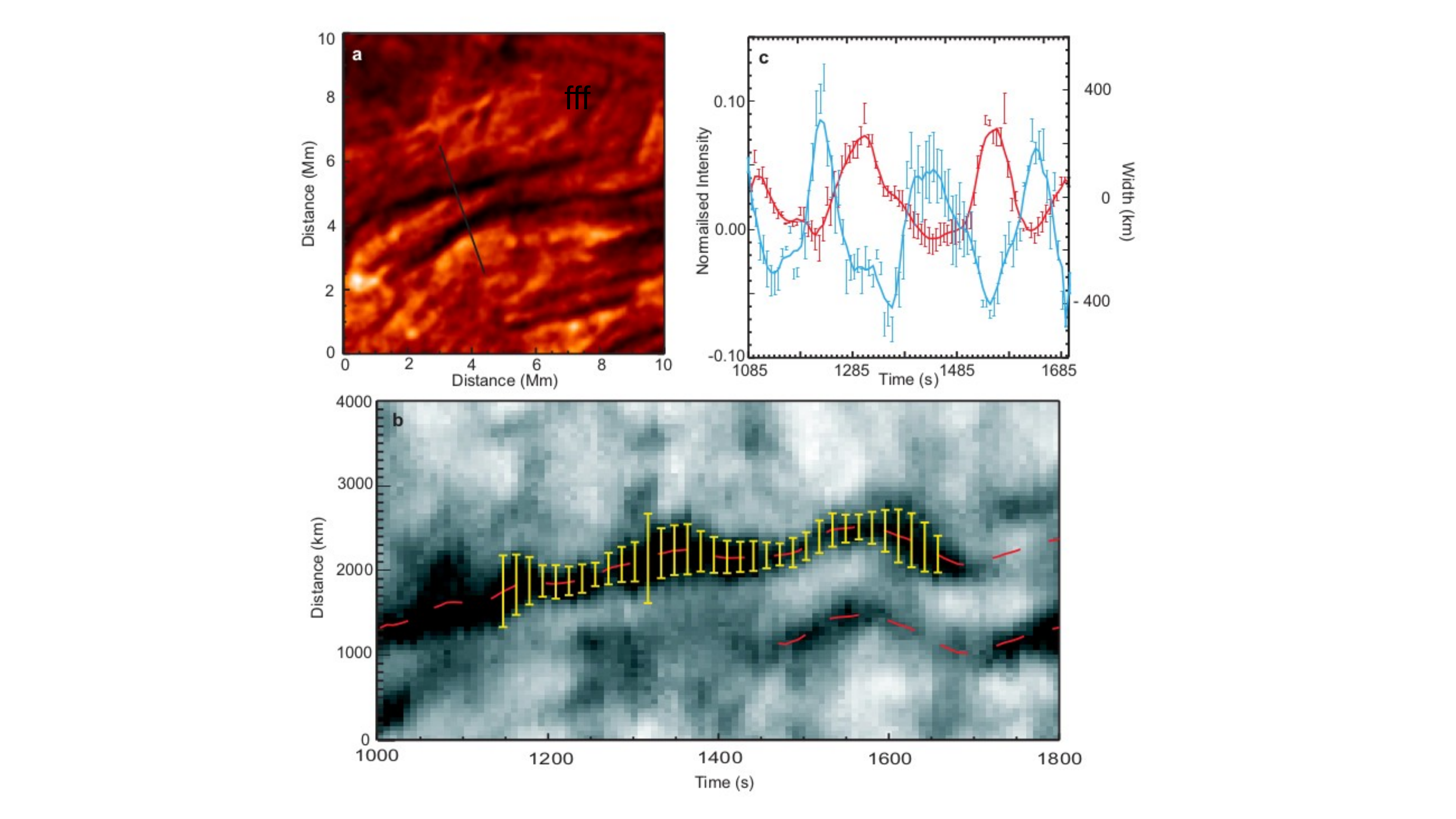}
%\centering {\includegraphics[width=1.6\textwidth]{f5}} 
\caption{Simultaneous wave modes in a  chromospheric magnetic flux tube using ROSA H$_{\alpha}$ observations. Time-distance map (`b') exhibits the displacement of the tube's axis (red-dotted line) as a whole (kink waves), while its cross-section variation (sausage wave) is depicted by the yellow bars measuring the variation of tube's width. All measurement is done by \citet{2012NatCo...3.1315M} from where this  figure is adopted and displayed here. Kink wave respectively have period, upward propagation velocity, amplitude, and energy flux as $\sim$232 s, $\sim$71 km s$^{-1}$, $\sim$5 km s$^{-1}$, 4300 W m$^{-2}$. While, the same for the observed sausage wave are given respectively as $\sim$197 s, $\sim$67 km s$^{-1}$, $\sim$1-2 km s$^{-1}$, 11700 W m$^{-2}$} \label{f5}
\end{figure*}
%%%%%%%%%%%%%%%%%%%%%%%%%%%%%%%%%%%%%%%%%%%%%%%%%%%%%%%%%%%% Figure 5%%%%%%%%%%%%%%%%%%%%%%%%%%%%%%%%%%%%%%%%%%%%%%%%%%%%%%%%%%%%%%%%%%%%   
In the chromosphere, magnetic structures play a major role in guiding MHD waves. They are thus the prime structures in which waves are detected. In particular, Fig.~\ref{f5} displays the results of \citet{2012NatCo...3.1315M} who have observed the simultaneous presence of fast kink and sausage waves in mottles and fibrils using ROSA H$_{\alpha}$ observations, and found that they carry an average energy of respectively 4300 W m$^{-2}$, and 11700 W m$^{-2}$, once again stressing the large amount of energy present in the lower atmosphere.\par

\begin{figure*} 
\hspace{-4.0cm}
\includegraphics[width=1.6\textwidth]{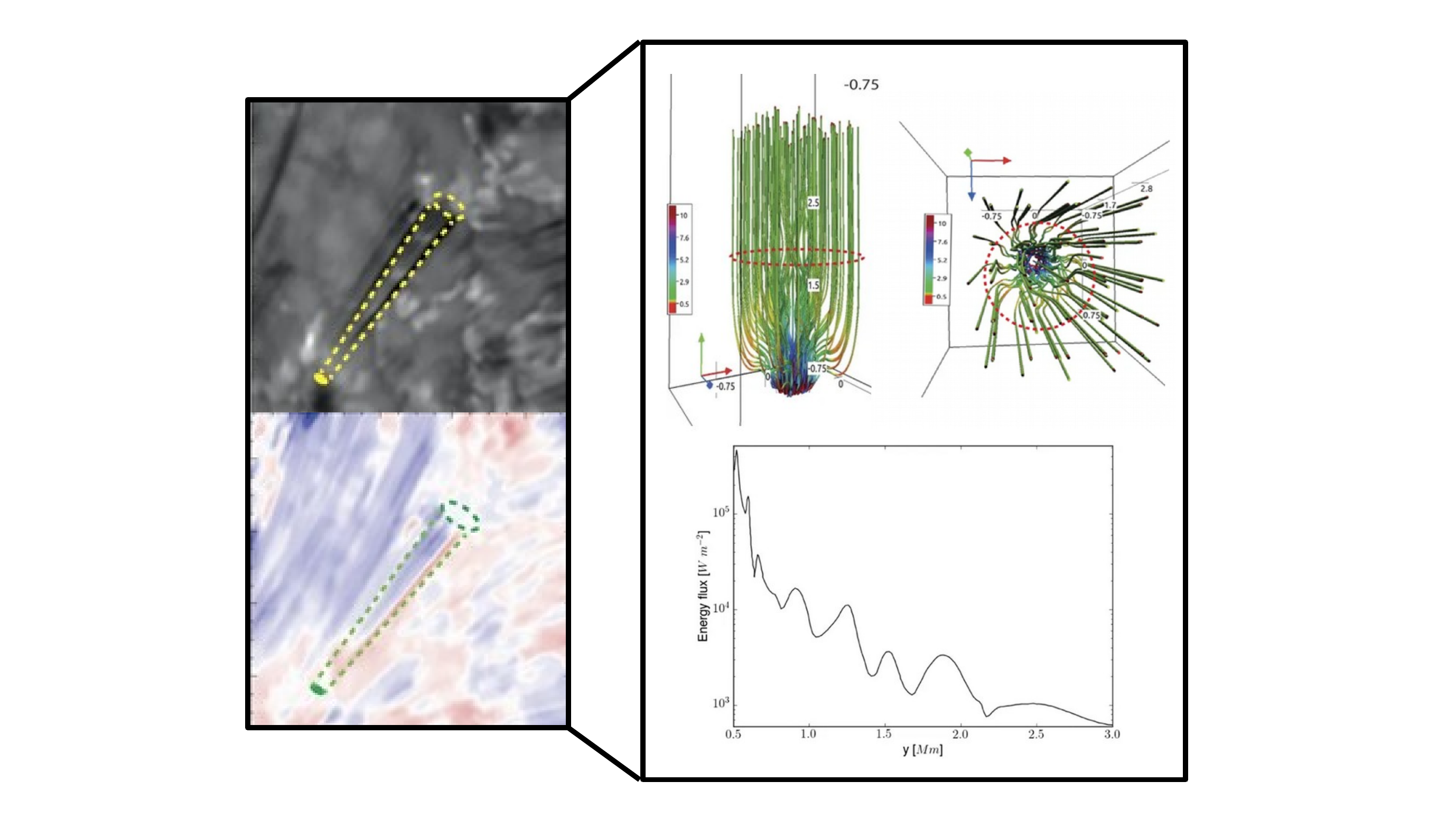}
%\centering {\includegraphics[width=1.6\textwidth]{f4}} 
\caption{The spicular-type structure (top panel of the left-box) shows periodic reversal of the velocity (snapshot shown in the bottom-panel of the left-box) indicating the presence of torsional Alfv\'en wave. The numerical simulation (top-panel of the right-box) exhibits similar torsional motions in the model fluxtube, that carry substantial energy (bottom-panel of the right-box) through the solar chromosphere, TR, and corona \citep[Credit:][]{2017NatSR...743147S}.} \label{f4}
\end{figure*}

Going from the chromosphere to the transition region, spicules and their TR counterparts show ample evidence of wave dynamics. 
Observations of swaying motions in spicules with high-resolution imaging instruments such as \textit{Hinode}/SOT and IRIS have long been attributed to propagating Alfv\'enic waves \citep{DePontieu_2007Sci...318.1574D,2014Sci...346A.315T}. The amplitudes of the swaying motions are usually of the order of $10-25$~km~s$^{-1}$, although a factor of 5 larger motions have also been observed \citep{2018ApJ...856...44A}. In general, 1 or 2 oscillations of the spicules can be captured before the structure disappears, and a wide range of reported periods of $100-500$~s and also high-frequency \citep[$20-50$~s][]{Okamoto_2011ApJ...736L..24O}. \citet{2017NatSR...743147S} showed the ubiquitous presence of high frequency ($\approx$12–42 mHz) torsional motions in spicular-type structures in the chromosphere (Fig.~\ref{f4}, left box). Their numerical model showed that these observations resemble torsional Alfv\'en waves associated with high frequency drivers containing a huge amount of energy ($\approx$10$^{5}$ W m$^{-2}$) in the chromosphere (Fig.~\ref{f4}, right-panel). It is important to note, however, that the observational signatures of TWIKH rolls can also explain these observations \citep{2018ApJ...856...44A}. Even after partial reflection from the transition region, \citet{2017NatSR...743147S} found that a significant amount of energy ($\approx$10$^{3}$ W m$^{-2}$) is being transferred into the overlying corona, which is sufficient to compensate the coronal radiative losses (Fig.~\ref{f4}, right panel). The propagation speeds of waves in spicules are often difficult to measure, due to their short lifetimes, the combination of (upward/downward) propagating and standing waves \citep{Okamoto_2011ApJ...736L..24O} and also due to the rapidly increasing Alfv\'en speed at the observing heights. On average, speeds of $200-300$~km~s$^{-1}$  are reported in these works. \par

 %%%%%%%%%%%%%%%%%%%%%%%%%%%%%%%%%%%%%%%%%%%%%%%%%%%%%%%%%%%% Figure 4%%%%%%%%%%%%%%%%%%%%%%%%%%%%%%%%%%%%%%%%%%%%%%%%%%%%%%%%%%%%%%%%%%%% 
%%%%%%%%%%%%%%%%%%%%%%%%%%%%%%%%%%%%%%%%%%%%%%%%%%%%%%%%%%%% Figure 4%%%%%%%%%%%%%%%%%%%%%%%%%%%%%%%%%%%%%%%%%%%%%%%%%%%%%%%%%%%%%%%%%%%%   

The swaying motions of spicules suggest a transverse displacement of a waveguide and therefore such motions are often interpreted as kink waves. However, caution is required when only imaging information is available, since torsional Alfv\'en waves may also produce strand-like, swaying structures such as spicules \citep{2018ApJ...856...44A}. Moreover, additional to the swaying, torsional and longitudinal motions are observed in spicules \citep{DePontieu_2012ApJ...752L..12D,Sekse_2013ApJ...769...44S}, and are likely strongly connected to their generating mechanisms and to the nature of the wave. The nature of the Alfv\'enic wave (kink or torsional Alfv\'en wave) is important when trying to estimate the energy flux carried by the wave, since the collective or local nature distributes the wave energy differently across the waveguide \citep{vd2014}. Assuming a filling factor of 1 (a best case scenario), a typical spicule mass density of $8-16\times10^{-14}~$g~cm$^{-3}$ and based on the reported values, one can estimate the energy flux in swaying spicules as $0.1-3\times10^{3}~$W\,m$^{-2}$, with the strongest cases at $1-4\times10^{4}~$W\,m$^{-2}$. Usually, $3\%$ of the energy is assumed to enter the corona, leading to average energy flux values of $100$ W\,m$^{-2}$ available for coronal heating, which is sufficient for the quiet Sun and for the acceleration of the solar wind \citep{2014Sci...346A.315T}.

%---------------------------------------------------------------------------------------------------------------------------------------------
%------------Heating Function Assessment -----------------------------------------------------------------------------------------------------

\subsection{Heating function assessment by slow waves}
The presence of waves in the solar atmosphere may not only provide energy for heating the corona, but could also provide us with a tool to seismologically estimate the coronal heating function and the related dissipation coefficients, such as thermal conduction and viscosity \citep[see the review by][this issue]{wang2020}. Here we forego that, more importantly, even the most basic property of the magnetized coronal plasma such as the magnetic field strength needed for any magnetically based heating function estimate is difficult to determine directly, and in some cases can only be estimated from coronal seismology. However, we keep this subject for the review of \citet{nakariakov2020}, this issue.

The impact on the dynamics of MHD waves of thermodynamic activity of the corona, i.e. processes of its continuous cooling via the optically thin radiation and thermal conduction, and resupply of energy by some yet unknown heating mechanism, was recently investigated by \citet{2017ApJ...849...62N, 2019A&A...624A..96C, 2019A&A...628A.133K, 2020arXiv201003364K}. This allows for developing a new approach for a remote diagnostics of thermodynamic properties of the corona, including processes of its cooling and heating, by coronal seismology. For example, additional restrictions on the coronal heating mechanism can be obtained via accounting for perturbations of the thermal equilibrium of the corona by compressive, e.g. slow magnetoacoustic, waves. Indeed, assuming the plasma heating and cooling processes are some different functions of the plasma thermodynamic parameters, i.e. density and temperature, and potentially of the magnetic field too, both of them can be perturbed by the waves. Such a wave-caused destabilisation of the initial thermal equilibrium leads to the onset of a \emph{heating/cooling misbalance} acting as an additional natural mechanism for the energy exchange between the plasma and the wave.

\citet{2019A&A...628A.133K} showed that it is convenient to use specific thermal misbalance time scales $\tau_1$ and $\tau_2$ connected to the rates of change of the net energy gain $H(\rho,T)$, and loss $L(\rho,T)$ including radiative cooling and field-aligned thermal conduction, through the function $Q(\rho,T)=L-H$ with the plasma density $\rho$ and temperature $T$, $Q_{T} \equiv \left( \partial Q  / \partial T \right)_{\rho}$ and $Q_\mathrm{\rho}\equiv(\partial Q/\partial \rho)_T $, as $\tau_{1}={\gamma C_\mathrm{V}}/\left[{Q_{T}-(\rho_0/T_0)Q_{\rho}}\right]$ and $\tau_{2}={C_\mathrm{V}}/{Q_{T}}$ with $C_\mathrm{V}$ being the specific heat capacity. With this,
%the energy equation, perturbed by a linear compressive wave, takes the form
%\begin{equation}\label{eq:energy_lin}
%\frac{\partial {T}}{\partial t} - (\gamma-1)\frac{T_0}{\rho_0}\frac{\partial {\rho}}{\partial t}=\frac{\kappa}{\rho_0 C_\mathrm{V}}\frac{\partial^2{T}}{\partial z^2} - \frac{{T}}{\tau_2}-\left(\frac{1}{\tau_2}-\frac{\gamma}{\tau_1}\right)\frac{T_0}{\rho_0}{\rho},
%\end{equation}
%where $\kappa$ is the field-aligned thermal conductivity. Considering the magnetic field to be infinitely stiff, so that the slow waves propagate strictly along the field, and using energy equation (\ref{eq:energy_lin}),
the dispersion relation for linear slow waves describes the evolution of two acoustic and one thermal (entropy-related) modes.
The 3rd order dispersion relation can be solved numerically for the real, $\omega_\mathrm{R}$, and imaginary, $\omega_\mathrm{I}$, parts of a complex frequency $\omega$, while allowing the value of $\omega_\mathrm{I}$ to be comparable to $\omega_\mathrm{R}$. The \emph{q-factor} (also known as quality factor) is computed as $q=\tau_\mathrm{D}/P$, using the oscillation period $P=2\pi\omega_\mathrm{R}^{-1}$ and damping time $\tau_\mathrm{D}=\omega_\mathrm{I}^{-1}$.

Using the radiative cooling $L(\rho,T)$ from the CHIANTI atomic database \citep{1997A&AS..125..149D, 2019ApJS..241...22D}, and a parametrised heating function $H(\rho,T)\propto \rho^aT^b$, \citet{2019A&A...628A.133K} compute numerically the oscillation quality factor $q$ as a function of the heating indices $a$ and $b$. 
\begin{figure}
	\centering
	\includegraphics[width=0.7\textwidth]{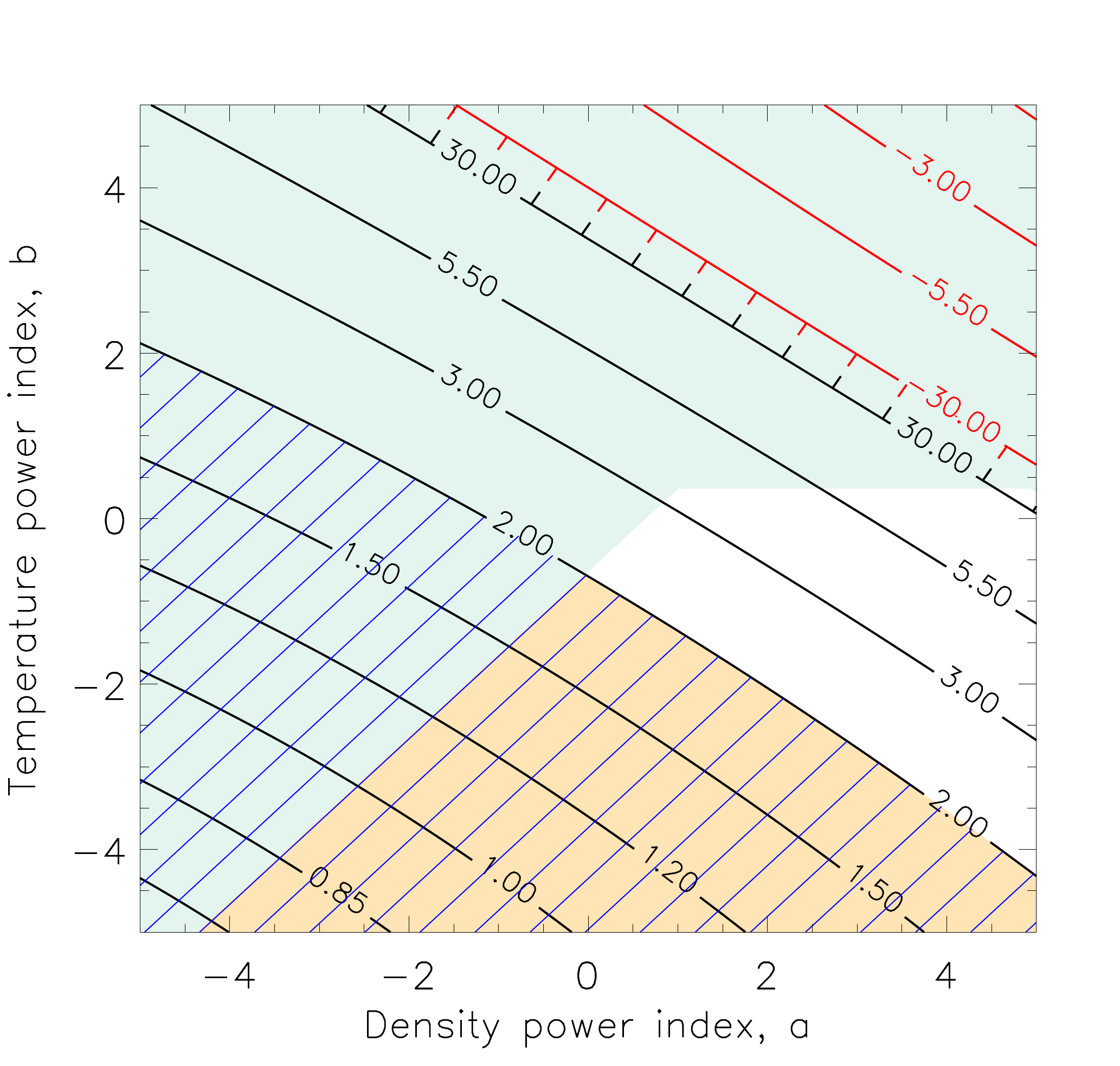} 
	\caption{Dependence of the oscillation quality factor $q$ on the density and temperature power indices used for parametrising the heating function as $H(\rho,T)\propto \rho^aT^b$, for the following values of the equilibrium parameters of the plasma: $T_0= 6.3\times 10^6\,\mathrm{K}$, $\rho_0 = 10^{-11}\,\mathrm{kg\,m}^{-3}$, loop length $L=180\times 10^6\,\mathrm{m}$, Spitzer conductivity $\kappa=10^{-11}T_0^{5/2}\,\mathrm{W\,m}^{-1}\,\mathrm{K}^{-1}$, mean particle mass $m=0.6\times1.67\times10^{-27}\,\mathrm{kg}$, $k_\mathrm{B}=1.38\times10^{-23}\,\mathrm{m}^2\,\mathrm{kg}\,\mathrm{s}^{-1}\,\mathrm{K}^{-1}$, and $\gamma =5/3$. The black and red contours show the regimes of damping (positive $q$) and amplification (negative $q$), respectively. The blue-shaded region shows values of $a$ and $b$ where the time scales of the heating/cooling misbalance $\tau_1$ and $\tau_2$ become negative, for which other thermal instabilities may occur \citep[see][]{1965ApJ...142..531F}. The hatched region shows values of $q<2$ for slow-mode oscillations detected in observations \citep{2019ApJ...874L...1N}. The yellow-shaded region outlines heating models which are stable to thermal instability and for which $q < 2$. Figure modified from \citet{2019A&A...628A.133K}.
	}
	\label{fig:quality_slow}
\end{figure}
The results are displayed in Fig.~\ref{fig:quality_slow} for different values of the heating power indices $a$ and $b$ and for the equilibrium parameters corresponding to hot and dense post-flare coronal loops.
Direct comparison of the values of $q$ shown in Fig.~\ref{fig:quality_slow} with those usually seen in observations of slow magnetoacosutic oscillations of coronal loops \citep[see e.g.][for the most recent statistical survey]{2019ApJ...874L...1N} by e.g. SOHO/SUMER and Yohkoh/BCS, shows that the heating models with $a$ and $b$ approximately delineated by a triangle with vertices $(-3.5,-4)$, $(0,-0.5)$, and $(5,-4)$ in Fig.~\ref{fig:quality_slow}, excluding the regions of the thermal mode instability, could be responsible for the observed rapid damping of slow waves in the corona \citep{2011SSRv..158..397W}. Thus, the proof-of-concept of \citet{2019A&A...628A.133K} show that slow waves can indeed serve to put limits on (power indices of) the coronal heating function. 

%There is currently a need for generalising the results shown in Fig.~\ref{fig:quality_slow} for the effects of a local obliquity of slow waves, i.e. for non-zero values of plasma $\beta$. In particular, \citet{2017ApJ...849...62N} developed a nonlinear theory of slow magnetoacoustic waves with the effect of the thermal misbalance, using the thin flux tube approximation and accounting for the potential dependence of the heating function on the magnetic field. However, the analysis was performed in the limit of weak non-adiabaticity and without a quantitative parametrisation of the heating model.

The discussed misbalance between heating and cooling processes in plasma can cause additional phase shifts between density and temperature perturbations in slow waves, thus affecting, for example, estimates of the effective polytropic index. In particular, \citet{2019PhPl...26h2113Z} demonstrated analytically that the polytropic index, a coefficient linking the slow wave phase speed with the plasma temperature, can vary non-monotonically with temperature due to the effect of the thermal misbalance, so that it deviates from the adiabatic value 5/3 to 1.4-3.2. This can be a natural cause for higher values of the polytropic index at hotter plasma temperatures observed by \citet{2018ApJ...868..149K}, that cannot be explained by the classical Spitzer thermal conduction. Implication of these theoretical results for probing the corona is a promising future research avenue.

%---------------------------------------------------------------------------------------------------------------------------------------------

%================================================================================================================================================

\section{Models}

\subsection{Phase mixing models}
\begin{figure}
	\includegraphics[width=.5\linewidth]{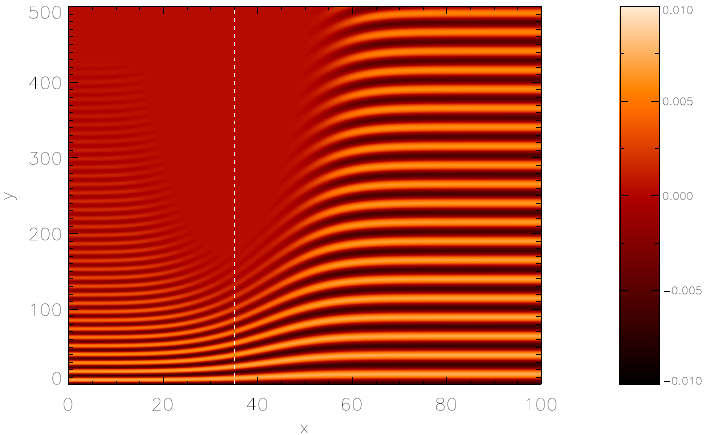}
	\caption{Snapshot of Alfv\'en wave velocity component out of the plane, for a periodic driver on the bottom boundary. The figure displays the turning of the wave fronts due to phase mixing in the middle region with the largest variation of the Alfv\'en speed, and the associated phase mixing damping in the region where the turning is the strongest, is indicated with the vertical white dashed line. Figure taken from \citet{2011A&A...527A.149M}.}
	\label{fig:phasemixing}
\end{figure}

As already pointed out by \citet{cowling1953} and \citet{piddington1956}, classical resistive or viscous dissipation of the Alfv\'en wave energy is not efficient in a homogeneous magnetized plasma and this slow dissipation rate remains a key obstacle in more advanced present day wave-based heating models \citep[see e.g.][]{arregui2015,cargill2016}. The main problem is to rapidly transfer the wave energy from large to small length scales, where (classical) dissipation is effective. The concept of phase mixing as described by \cite{heyvaerts1983} aims to enhance this dissipation rate through the creation of small length scales in an inhomogeneous medium. Phase mixing occurs when a local gradient in the Alfv\'en speed is present; as waves propagate along field lines with different Alfv\'en speeds, the wave front turns and the waves become increasingly out of phase. This is shown graphically in Fig.~\ref{fig:phasemixing}, displaying results of \citet{2011A&A...527A.149M}. This process generates increasingly smaller length scales (across the background magnetic field) and hence, more efficient dissipation. Although resonant absorption and phase mixing are often discussed as individual mechanisms, there is a natural link between these mechanisms, as both rely on the presence of a variation in the local Alfv\'en speed profile \citep{soler2015}. Hence, the small-scale oscillations in the resonant layer will naturally undergo phase mixing due to local variation in the Alfv\'en speed \citep[see e.g.][]{ruderman1997,ruderman1997b}. However, the damping caused by phase mixing depends on the actual value of the resistivity, while the damping of resonant absorption does not (as resonant absorption itself is an ideal process). An extensive body of literature on both resonant absorption and phase mixing exist, for which we refer the interested reader to reviews by e.g. \citet{Asc04,goedbloed2004,goossens2011}. 

Most theoretical (including computational) studies of phase mixing start from an equilibrium setup where a pre-existing profile in the Alfv\'en speed is present. Most often, this setup consists of a uniform background magnetic field, where a gradient in the density is balanced by a temperature variation to maintain pressure balance. Waves are then considered as perturbations of this equilibrium or injected through boundary driving. The process of phase mixing will lead to the most efficient dissipation of the Alfv\'en waves (and hence strongest heating) where the gradient in the density (Alfv\'en speed) profile is steepest. 

The self-consistency of heating by phase mixing of Alfv\'en waves was investigated in more detail by \citet{cargill2016} by analysing the evolution of the density profile. These authors showed that although phase mixing increases the efficiency of the wave energy dissipation where gradients in the local Alfv\'en speed occur, the resulting heating in this basic model cannot self-consistently sustain the required density profile in closed loops. As the local density in closed loops is related to the magnitude of the heating \citep[see e.g.][]{klimchuk2006,reale2014}, the phase mixing heating profile is not consistent with the heating profile to sustain the density profile. \citet{cargill2016} also investigated whether feedback of the heating on the density profile through evaporation would be able to modify the local density \citep[see also][]{ofman1998}. Although some local structuring of the density profile occurred, it only happened on timescales longer than the cooling and draining timescales and hence, does not help address the efficiency problem of wave-heating. In addition, the authors point out that transport coefficients need to be substantially enhanced to obtain effective heating in the first place. Using MHD simulations including thermal conduction and optically thin radiation, \citet{vandamme2020} for the first time modelled the feedback process through evaporation entirely self-consistently (i.e. without the use of scaling laws) and also found that Alfv\'en wave phase mixing only leads to modest heating in the shell regions of the loop, where the mass increase through evaporation is not sufficient to modify the phase mixing process.

If the imposed density structure is not compatible with the heating profile resulting from phase mixing, how then is this assumed equilibrium structure supported? Is an alternative heating mechanism present that is compatible with the density structuring? And if so, does that immediately imply that the wave-based heating (through phase mixing) is comparatively small?

In addition to requiring the presence of a variation in the local Alfv\'en speed profile, the phase mixing model as it was originally introduced by \cite{heyvaerts1983} implies the presence of an ignorable coordinate, a setup which might not be representative of the highly inhomogeneous solar atmosphere \citep{parker1991,1995JGR...10023413O}, although approximately similar  conditions may occur in coronal holes. 

\cite{2017A&A...601A.107P} investigated the heating by phase mixing in a 3D numerical model of a boundary-driven flux tube. The flux tube is modelled as a cylindrical density enhancement in a uniform magnetic field and transverse displacements of the footpoint of the cylinder generate kink modes which, as they propagate along the flux tube, couple to azimuthally polarised Alfv\'en modes in the boundary shell of the cylindrical flux tube. Due to the density gradient in the inhomogeneous boundary layer, the Alfv\'en waves phase mix but, even using (excessively) large values of magnetic resistivity and large-amplitude footpoint driving, the heating due to phase mixing was found to be insufficient to be relevant for coronal heating (i.e. to balance the expected losses through radiation and conduction), a conclusion similar to the remark made by \citet{cargill2016} about the need to substantially enhance transport coefficients.  By varying parameters such as the length of the non-uniform layer, the density structure, and the persistence of the driver, 
\cite{2017A&A...601A.107P} find that phase mixing of these propagating waves leads to temperature increases of the order of 10$^5$K or less, a figure that is in agreement with the analytical estimate by \cite{terradas2018b}, even though these were for standing waves. \par

When simulations are performed using 3 observed coronal loop oscillation harmonics as input, \cite{pagano2018} found that the presence of these multiple harmonics causes drifting of the location of the heating. Still, the mechanism did not seem to provide enough energy to maintain the full thermal structure constrained by the observed coronal properties, and the multiple harmonics inhibited the formation of small scales. \citet{pagano2019} included an observed spectrum of transverse waves, using a boundary driver which consists of a series of 1000 superimposed random pulses was used drawn from a reconstructed spectrum.  The results again indicate that it is unlikely that phase mixing of Alfv\'en waves generated by the observed power spectrum heats coronal loops, although, the waves could be important in the generation of small scales. 

\subsection{Alfv\'en wave heating models}

\subsubsection{Alfv\'en wave induced shock heating}

Alfv\'en waves have long been particularly attractive coronal heating candidates due to their ability to carry large amounts of energy throughout the solar atmosphere \citep{1947MNRAS.107..211A,Uchida_1974SoPh...35..451U} and can potentially also lead to solar wind acceleration \citep{1990JGR....9514873H,2010LRSP....7....4O}. The Poynting flux upward from convective motions in magnetic concentrations at photospheric level is expected to be on the order of $10^{6}$~W\,m$^{-2}$ \citep{Parnell_DeMoortel_2012RSPTA.370.3217P}, although the strong magnetic expansion in the upper layers leads to an effective Poynting flux of $10^{4}$~W\,m$^{-2}$ into the chromosphere. The ion-neutral friction due to the partial ionisation of the plasma in the chromosphere strongly damps the Alfv\'en waves, in particular the high-frequency spectrum. Combined with reflection, these effects lead to only a $1-3$~\% effective transmission rate into the corona \citep{Soler_2017ApJ...840...20S,2019ApJ...871....3S}, similar to estimated footpoint leakage of coronal waves \citep{depontieu2001}. Overall, the energy budget from torsional Alfv\'en waves generated in the photosphere is estimated to be on the order of $10^{2}$~W\,m$^{-2}$ in the corona, which is just enough for the quiet Sun or coronal hole. The double mode conversion process from p-mode waves into Alfv\'enic waves in the chromosphere and transition region is expected to contribute similar energy rates \citep{Morton_2019NatAs.tmp..196M}.\\
Various models for Alfv\'en wave propagation in the solar atmosphere have a different nature depending on the region and wave guide under consideration. The geometry of the waves has been selected to be either planar \citep[a.k.a linear, e.g.,][]{2005ApJ...632L..49S,2010A&A...518A..37M} or azimuthal  \citep[a.k.a. circular, e.g.,][]{1996PhPl....3...10Z,2010A&A...517A..29V,2017SoPh..292...31W}. For instance Alfv\'en wave propagation of waves with linear polarisation may be more appropriate in the relatively diffuse coronal holes, while the circular polarisation may be more applicable for the propagation in structures showing strong density contrast, such as solar jets, spicules, tornadoes, and loops \citep{2011A&A...526A..80V}. \par

Alfv\'en waves need an efficient dissipation mechanism in order to play a dominant role in coronal heating. A first successful dissipative model was based on nonlinear mode conversion of these waves into compressive modes (evolving into shocks) due to the flux tube expansion (and associated centrifugal force) and the ponderomotive force \citep{Hollweg_1982SoPh...75...35H,1996ApJ...465..451L,1998JGR...10323677O,2010ApJ...712..494A}. This idea has persisted through the decades thanks to the accompanying ability to generate spicule-like excursions of material into the corona \citep{Kudoh_1999ApJ...514..493K,2010ApJ...710.1857M,2016ApJ...817...94A,2016ApJ...829...80B} and accelerate the solar wind \citep{2014MNRAS.440..971M,2006JGRA..111.6101S}. Furthermore, this model has demonstrated that waves can lead to small nanoflare-like intensity bursts during dissipation \citep{moriyasu2004,antolin2008}, thus placing caution when interpreting observations of these events or associating the nanoflare term solely to reconnection-based models. Although highly self-consistent in its ability to explain various features of the solar atmosphere, this model relies on the presence of density fluctuations, and sufficiently large transverse wave fluctuations (compared to the local Alfv\'en speed) for the relevant nonlinear effects to become important. A natural question is whether this matches observations and the answer yet unknown. Parker Solar Probe does indicate, however, that the density fluctuations in the open corona is far greater than previously thought \citep{2019Natur.576..237B}.
 
\begin{figure}
	\includegraphics[width=.5\linewidth]{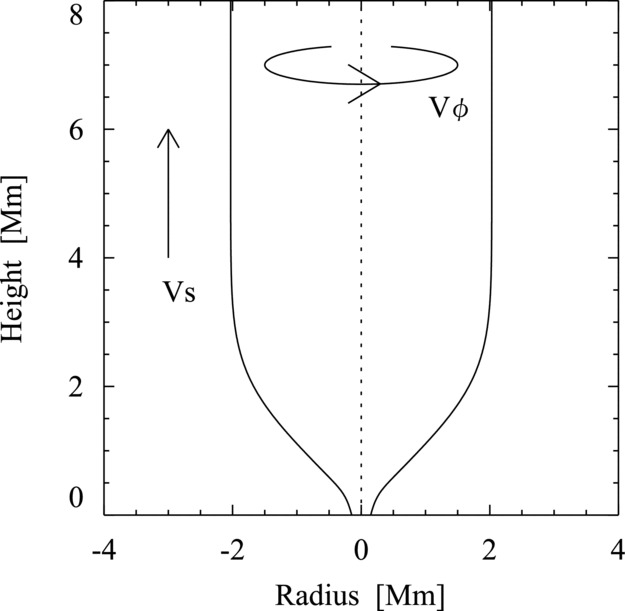}\ \includegraphics[width=.5\linewidth]{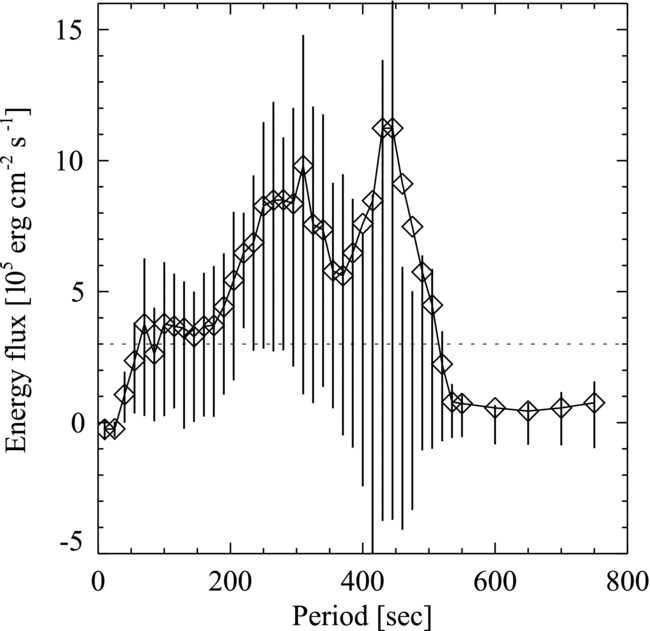}
	\caption{Left: Sketch on the configuration of Alfv\'en wave heating models. Right: Alfv\'en wave energy flux as a function of period. The figures are taken from \citet{2010ApJ...710.1857M}.}
	\label{fig:alfven}
\end{figure}

A well established model is that of \citet{2005ApJ...632L..49S} who describe the propagation of a low frequency Alfv\'en wave in coronal holes from the photosphere to an altitude of 0.3AU. The granular motion of the photosphere results in steady transverse motions perturbing the magnetic field lines and exciting Alfv\'en waves \citep{2005ApJS..156..265C}. The 1D model provided by \citet{2005ApJ...632L..49S} resembles a superradial open magnetic flux tube conserving magnetic flux, where the expansion is a two step superradial function with the use of two expansion factors. Their choice for a two step superradial function was due to the height dependence of the magnetic field strength reconstructed by \citet{2005Sci...308..519T}. They included the effect of field line curvature, radiative cooling and thermal conduction in the MHD equations. The solutions to these MHD equations enabled \citet{2005ApJ...632L..49S} to compute the variations of the radial and tangential speeds together with the temperature and density as a function of altitude (see Fig.~\ref{fig:alfven} for a sketch of the configuration). The results proved adequate for atmospheric heating due to two effects; the first stage of heating is due to the dissipation of low frequency Alfv\'en waves mainly heating the inner solar atmosphere, while the second stage of heating is due to the induction of compressive perturbations (especially slow waves) due to the nonlinear efects connected with the Alfv\'en wave propagation that causes wave steepening ending up in shock formation that mainly contributes towards coronal heating. The nonlinear effects connected with outward propagating Alfv\'en waves that experience shocks contributes in two ways; the first aspect is the heating due to the increase of wave amplitudes that is confirmed by the non-thermal broadening of the emission lines, the second aspect is the creation of shocks which rapidly dissipates the waves. The contribution of the second aspect  towards heating is greater than the first aspect \citep{2000A&A...353..741N}. As a matter of fact, the efficiency of damping and hence heating due to low frequency Alfv\'en waves also depends on the activity of the region. In particular, the damping of surface Alfv\'en waves occurs on a shorter scale in active regions compared to quiet solar regions. This means that in quiet regions, the surface Alfv\'en wave is able to contribute towards heating the corona at higher altitudes \citep{2009ApJ...703..179E}. In addition, the period of the Alfv\'en wave itself is key in determining its contribution to coronal heating. This statement is backed by the simulations of \citet{2010ApJ...710.1857M} where nonlinear Alfv\'en waves were driven by photospheric convection towards the transition region in the presence of gravity and empirical chromospheric cooling. 
Their model showed that the region bounded between the photosphere and the transition region acts as a resonant cavity, a fact that has been recently observed above sunspots for the first time \citep{2020NatAs...4..220J}. It can be readily noticed from the right panel of Fig.~\ref{fig:alfven} that waves with periods between $100$ s to $500$ s are able to transport flux to the corona, and that the maximum energy flux is carried when the period of the Alfv\'en wave is around $400$s coinciding with the resonant frequencies in the cavity \citep{2010ApJ...710.1857M}. \par
Coronal resonances of Alfv\'en wave propagating in solar loops have also been reported in the model proposed by \citet{2010ApJ...712..494A}. The loop is a gravitationally stratified semitorus, with magnetic field expansion, similar to the left panel of Fig.~\ref{fig:alfven}. The advantage of this model is that it sheds light on the efficiency of heating taking into account the loop geometry in which Alfv\'en waves propagate. This enabled \citet{2010ApJ...712..494A} to state that Alfv\'en waves provide efficient and uniform heating when propagating in thick loops where the area expansion between the photosphere and corona is greater than 2.5 orders of magnitude with lengths at least as long as 80 Mm. Since in the solar quiet regions the loops are both longer and wider compared to loops in active regions, the efficiency of Alfv\'en wave heating in quiet solar regions is more pronounced compared to active regions. Regarding the period of the waves, \citet{2010ApJ...712..494A} concluded that the shocks connected with long-period waves increase the average temperature of the corona, while shocks connected with short-period waves are unable to further heat the corona despite being more numerous \citep{2006JGRA..111.6101S}. The short-period waves are only just strong enough to maintain the coronal temperature. Heating by the resonance cavity with a monochromatic driver matching the eigenfrequency of the loop was found to lead to temperatures close to 5 MK, but the loops were not in a state of thermal equilibrium.\par 

Regarding the contribution of the compressive shocks towards coronal heating,  \citet{2016MNRAS.463..502M} stated that shock  compressive heating is very efficient below the altitude of 4 Mm, while above this height the incompressive heating due to direct dissipation of magnetic and velocity shear in Alfv\'en waves is dominant. Moreover, \citet{2014MNRAS.440..971M} showed that these chromospheric compressive shocks generate curved wedge shaped Alfv\'en waves, which could play a role in heating the higher layers. Thus, the back reaction of the induced compressive perturbations on the Alfv\'en wave results in Alfv\'en wave shocks that also contribute towards coronal heating. The induced compressive perturbations are due to the nonlinear forces connected with Alfv\'en waves, namely the magnetic tension, centrifugal, and ponderomotive forces. It is worth noting that the ponderomotive coupling of Alfv\'en waves to slow modes creates shocks that dominates the heating due to resistive dissipation \citep{2016ApJ...817...94A,2016ApJ...829...80B}. The non-linear behaviour of Alfv\'en waves was described by  e.g. \citet{suzuki:hal-00302984,2011A&A...526A..80V,2012EP&S...64..201S,2017ApJ...844..148V}. In particular, \citet{2012A&A...544A.127V} implemented the second order thin flux tube approximation \citep{1996PhPl....3...10Z} and studied the parallel nonlinear cascade of torsional and shear Alfv\'en waves in open magnetic fields. They showed that the shock formation for shear Alfv\'en waves comes into play earlier than torsional waves in the lower solar atmopshere. \\
Another mechanism playing a role in chromospheric heating is Ohmic diffusion and ion-neutral collisions which strongly affects torsional Alfv\'en waves during their propagation in the chromosphere \citep{2019ApJ...871....3S,2020AA...635A..28W} and leads to heating. In any case, the chromosphere plays a crucial role for coronal heating, because of the resulting evaporation of material due to coronal heating and thus providing the mass source. The localised heating could result in thread structuring in the corona \citep{2008A&A...478..921C}, which is in contradiction with the recent results of \citet{cargill2016,vandamme2020}.

\subsubsection{Alfv\'en wave turbulence heating}
\begin{figure}
	\includegraphics[width=\linewidth]{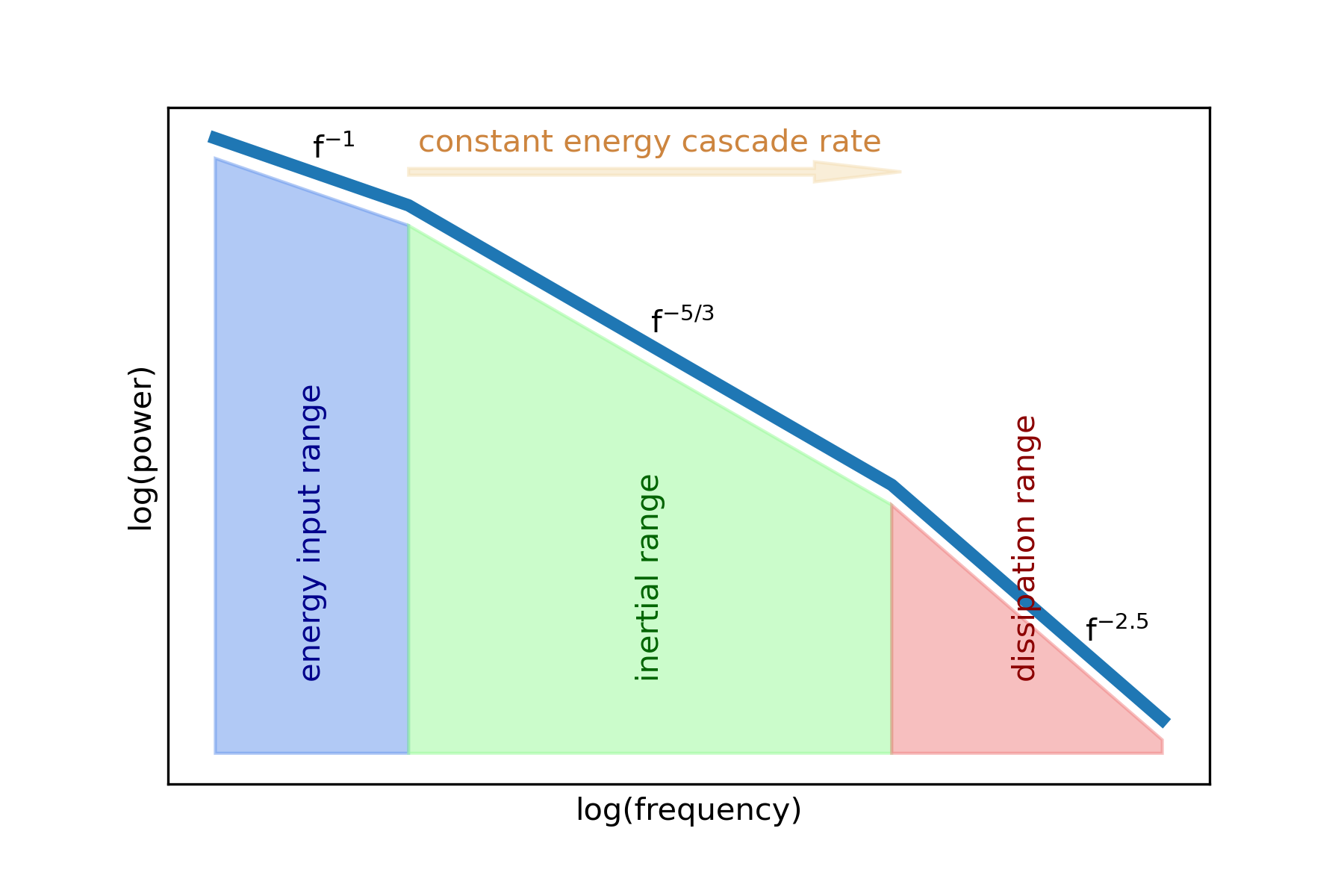}
	\caption{A schematic graph of the power law behaviour of turbulent fluctuations in the solar atmosphere and wind. The left, blue range shows the energy input range with a power law index of -1. The middle, green range shows the inertial range with a power law index of -5/3, where the energy cascade rate is independent of the scale. The right, red range shows the dissipation range, where the ever steeper power law slope depends on the specific dissipation process.} 
	\label{fig:powerlaw}
\end{figure}

Turbulent heating models have gained popularity recently in the context of coronal heating \citep[see the review by][of this mechanism in the solar wind context]{2013LRSP...10....2B}. However, there is still no direct evidence that the corona is turbulent. This might change in the near future thanks to the Parker Solar Probe, which could confirm turbulence at heights around or below the Alfv\'en critical point. Still, the non-thermal broadening of the coronal spectral lines could already be indicative of turbulent fluctuations \citep{banerjee1998,2006SoPh..236..245S,hahn2013,hahn2014}. More indirect evidence is present in the CoMP observations through the measurement of a $1/f$ spectrum in closed loops \citep{Morton_2016ApJ...828...89M,Morton_2019NatAs.tmp..196M}. Turbulent heating models are best categorized as wave or alternative-current (AC) mechanisms in the limit of weak turbulence, as the turbulent energy cascade is thought to be generated by nonlinear (self) interactions of waves. The underlying idea of energy conversion is similar to other wave-based heating mechanisms: increasingly smaller length scales are created, until dissipation becomes important and converts kinetic and magnetic energy into heat. The main difference compared to other mechanisms is that in turbulence small scales are created nonlinearly (unlike in phase mixing, resonant absorption, etc.), producing an energy cascade  to smaller scales (although, in some scenarios inverse cascade can also take place). We differentiate three main scales or ranges in turbulence, as depicted schematically in Fig.~\ref{fig:powerlaw}. The largest scale is the energy containing scale, at which the forcing of the plasma takes place, e.g. the size of convective cells in the photosphere (depicted in blue in Fig.~\ref{fig:powerlaw}). The inertial range is an intermediate scale where the actual energy cascade is initiated. The dynamics are self-similar and independent of both the nature of forcing and dissipation (shown in green in Fig.~\ref{fig:powerlaw}), expected to  correspond to Kolmogorov power law indices of -5/3. Lastly, at scales where dissipative terms are on the order of advective terms (Reynolds and/or magnetic Reynolds numbers $\approx 1$), turbulence enters the dissipation range, where heat is generated  (assuming collisional dissipation), displayed with the red zone in Fig.~\ref{fig:powerlaw}. In this last range, the power law index depends on the particular dissipation mechanism (e.g. electron or proton dissipation) that is considered, growing ever steeper. \par

\begin{figure}[t]
  \centering
  \medskip
  \includegraphics[width=0.75\textwidth]{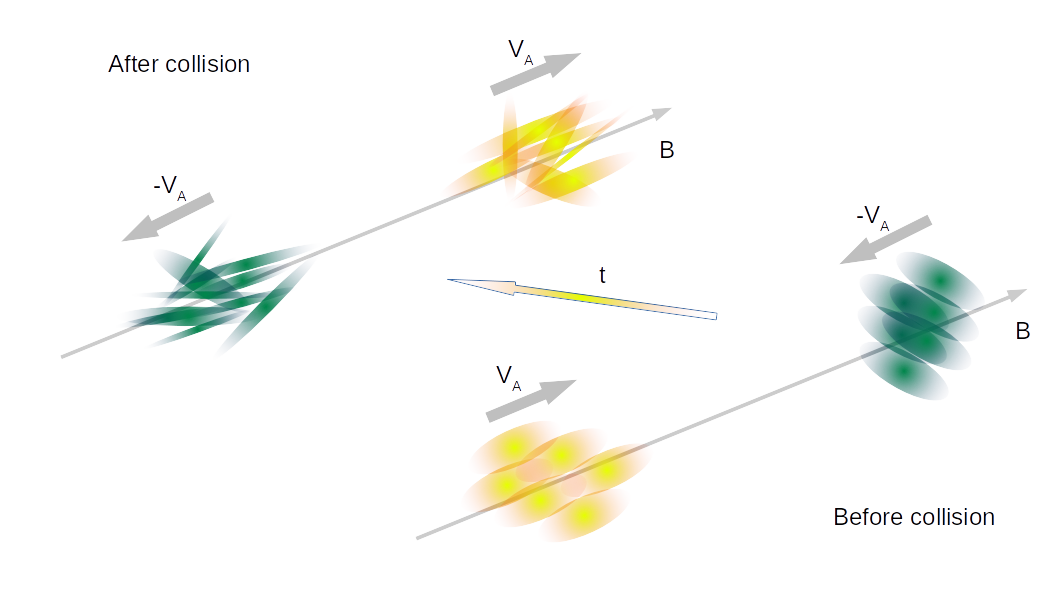}
  \caption{Illustrating the mutual deformation of two counter-propagating Alfv\'enic wave-packets along some background magnetic field $\mathbf{B}$. The undisturbed wave-packets must collectively vary along both perpendicular directions for nonlinear interaction to occur \citep{2013PhPl...20g2302H}, depicted by the different orientation of the ellipsoid making up the packets. These ellipsoids can be considered isosurfaces of velocity perturbations. After the collision (on the upper left), the wave-packets are mutually deformed, leading to ellipsoids scattered to higher wavenumbers, i.e. appearing thinner. } 
  \label{one}
\end{figure}

The most researched turbulence-generating wave interaction is that of counter-propagating pure Alfv\'en waves in the incompressible limit. In this scenario, the Alfv\'en wave wavefronts are deformed in successive collisions, leading to a cascade towards higher wavenumbers (see Figure~\ref{one}).
In the corona, counter-propagating transverse or Alfv\'enic waves can exist in both closed structures, such as coronal loops, and open structures, due to wave reflection. Wave reflection occurs when the Alfv\'en speed varies along the propagation direction, i.e. along the background magnetic field, linearly coupling the outward and inward-propagating Alfv\'en waves \citep{1980A&A....91..136L,2007JGRA..112.8102H}. Alfv\'en speed variations along the field exist in the corona because of gravitational stratification of the plasma, and the spherical/dipolar expansion of the magnetic field with height. In the turbulent heating models, the source of input energy is usually the convective flows in the photosphere, which generate waves and eventually turbulence by shuffling the magnetic field lines. There are numerous models of coronal heating based on turbulence for both coronal loops \citep{2011ApJ...736....3V,2016ApJ...832..180D,2017ApJ...849...46V}, and open structures \citep{2013ApJ...776..124P,2015ApJ...811..136W,2019JPlPh..85d9009C}, and also global models \citep{2014ApJ...782...81V,2017ApJ...845...98O}. The models for Alfv\'en wave heating of active regions \citep{2018JPhCS1100a2027V} show that temperatures reach maximally 2.5 MK, and that other heating mechanisms are needed to go beyond those temperatures.\\
The `local' models mostly use a reduced MHD treatment, which neglects density perturbations  (i.e., are incompressible) and transverse inhomogeneities in the plasma, among other simplifications. For the global models, Alfv\'en wave equations are employed in \citet{2014ApJ...782...81V}, containing heavily approximated terms for reflection and turbulent dissipation, and Reynolds-averaged mean flow equations with turbulent transport in \citet{2018ApJ...865...25U}. All these models conclude that a multi-million K corona can be maintained by turbulent heating, while balancing radiative cooling and conductive losses. A common discrepancy between turbulent heating models and observed properties of the corona is the excessive non-thermal line broadening that results from the models \citep{2016ApJ...820...63B}. \par

\begin{figure}[t]
  \centering
  \medskip
  \includegraphics[width=0.75\textwidth]{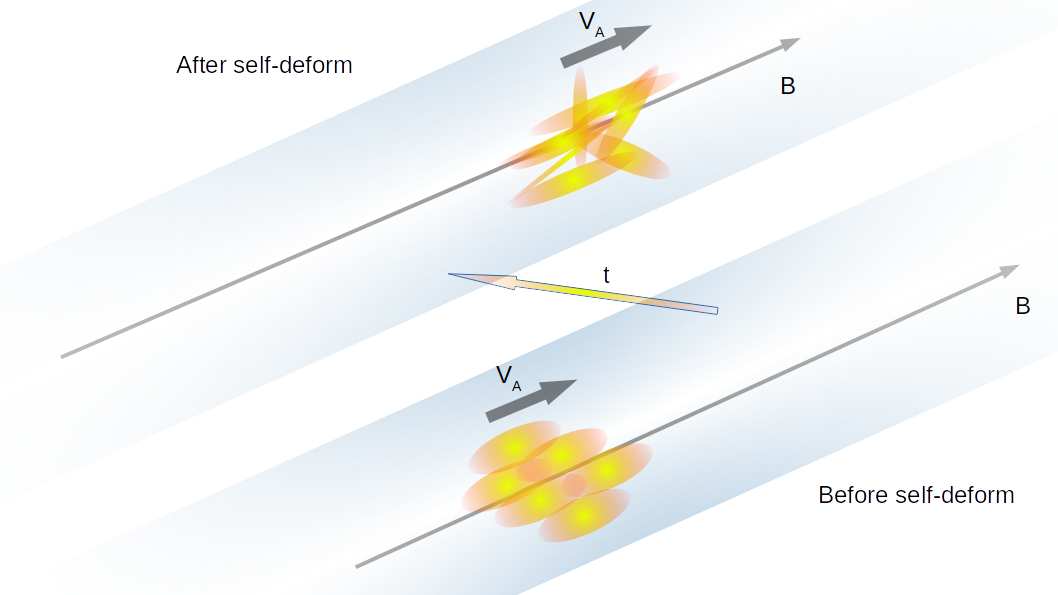}
  \caption{Same as in Figure~\ref{one}, but having only the wave-packet propagating parallel to the magnetic field. The background plasma is inhomogeneous across the field. This is causing the initial Alfv\'enic wave-package to self-deform nonlinearly as it propagates, leading to elongated ellipsoids, i.e. cascading to higher wavenumbers.} 
  \label{two}
\end{figure}

Recently, it was realized that the existence of counterpropagating waves is not the only way to generate turbulence \citep{2017NatSR...14820M,2019ApJ...882...50M}. In the presence of transverse inhomogeneities, propagating Alfv\'enic waves can nonlinearly self-deform, generating a cascade to smaller scales, called uniturbulence (see Figure~\ref{two}). 
This can constitute an additional channel towards a turbulent cascade and can potentially enhance the dissipation rate, leading to more heating. However, there are no models of coronal heating incorporating the effects of uniturbulence, as of yet. Still, forward modelling of simulations with uniturbulence \citep{pant2019} can elegantly and self-consistently explain the observed correlation between wave Doppler shift and spectral line broadening \citep[][as also explained in \citeauthor{nakariakov2020},\citeyear{nakariakov2020}]{mcintosh2012}.

Full 2.5D and 3D MHD models are now available \citep[e.g.][]{2012A&A...544L..20D,2019ApJ...880L...2S} that take into account both the nonlinear Alfv\'en wave heating via compressive effects, and the Alfv\'en wave turbulence, and initial results seem to point to the dominance of Alfv\'en wave turbulence over the compressive heating in the corona, while the opposite is true at the lower heights of the magnetic canopy \citep{2016MNRAS.463..502M,Matsumoto_2018MNRAS.476.3328M}. Since these models target quiet Sun conditions for the loops it is not yet clear whether the same holds for the denser and more dynamic active region loop conditions.

To date, all Alfv\'en wave heating models have been mostly successful at generating and sustaining a lower energy budget corona as in the quiet Sun or coronal holes, while they all seem to fall short of the $10^{4}$~W\,m$^{-2}$ average heating rate needed for active region coronal loops \citep[see chapter 6.1][for a more thorough discussion]{Hinode_10.1093/pasj/psz084}. In particular, the hot loops at the core of active regions with temperatures of 5~MK or more seem to be unexplained by wave models in general \citep{2018JPhCS1100a2027V}. Furthermore, these models seem to generate preferentially uniformly heated coronae in line with the RTV scaling law \citep{Rosner_1978ApJ...220..643R}, rather than footpoint heating. In turn, Alfv\'en wave heating models do not seem able to explain thermal non-equilibrium and the associated coronal rain \citep{Antolin_2010ApJ...716..154A}. Comparison of results from the Alfv\'en wave turbulence model with \textit{Hinode}/EIS observations of active region loops indicates that longitudinal flows are an important ingredient, suggesting that either compressive effects may become important or that a different mechanism is present \citep{2014ApJ...786...28A}. It is important to note, however, that the very demanding computational requirements needed in fully consistent wave models strongly limit the current predictions (see section~\ref{wh_conclusions})

\subsection{Kink-mode driven turbulent heating models}\label{sec:kinkkhi}
\begin{figure}
	\includegraphics[width=.7\linewidth]{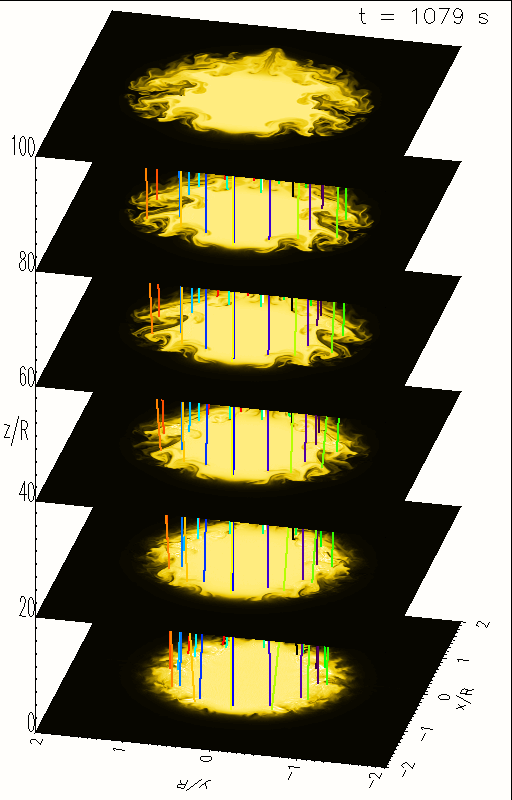}
	\caption{A tomography of TWIKH rolls. Cross-sections of density (in yellow) are shown at different locations along half the length of a prominence oscillating with a standing kink mode after roughly 2 periods. The coloured lines correspond to selected magnetic field lines traced from the bottom. The TWIKH rolls drag the field lines, generating twisting and strong changes in magnetic field pressure in a wide layer around the boundary. The simulation corresponds to the high resolution run discussed in \citet{antolin2015}. }
	\label{fig:KHI}
\end{figure}

\begin{figure}
	\includegraphics[width=\linewidth]{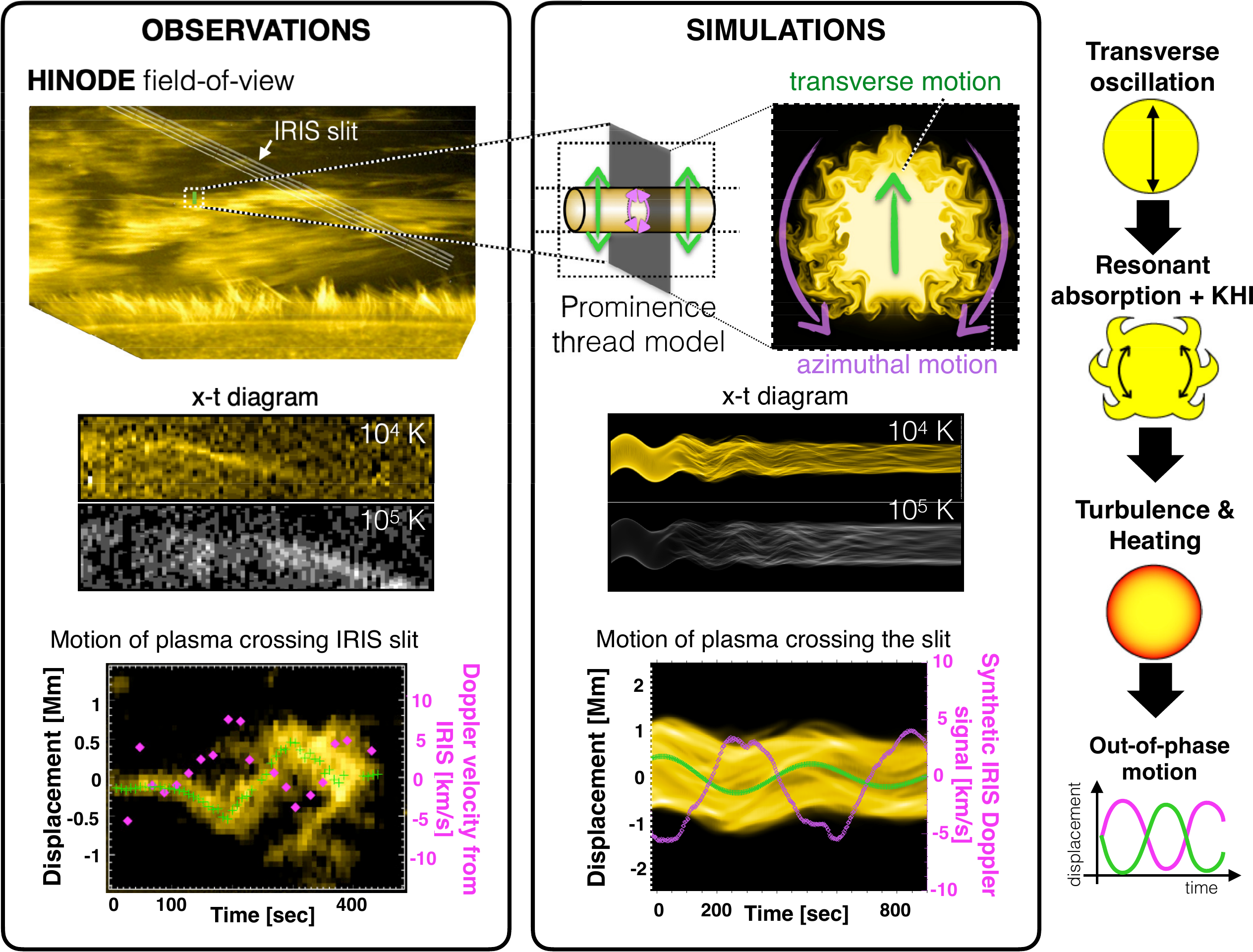}
	\caption{Evidence for resonant absorption and heating in a prominence. The left column provides a set of observables obtained with coordinated observations (top panel) between Hinode/SOT and IRIS of an active region prominence off-limb. (Left-middle) Prominence threads were seen oscillating transversely and fading in the Ca II H line of SOT ($10^4$ K) and subsequently appearing in the hotter SJI 1400 channel of IRIS (dominated by Si IV emission at $10^{4.8}$~K approximately). (Left-bottom) The prominence threads oscillate in the POS (yellow with green curve) out-of-phase with the Doppler velocity in Si IV captured by the IRIS slit (purple). The right column shows the forward modelling of a numerical simulation of a prominence thread with the observed characteristics. The thread oscillates transversely with a kink mode and triggers the KHI at the boundary (top-right). Resonant absorption is initially highly localised in the boundary, which fuels the KHI. TWIKH rolls are excited, whose dynamics reflect the resonant absorption and phase mixing dynamics but at a larger, observable scale. The TWIKH rolls mix the cold prominence plasma with the hot coronal surroundings and widen the boundary of the loop. A turbulent cascade is established which leads to wave dissipation and heating. As the prominence plasma is heated, its emission in the synthesised Ca II H line fades out and becomes stronger in the hotter Si IV line (right-middle). The resonant absorption and phase mixing dynamics become observable due to the KHI, leading to an out-of-phase relation between the POS motion of the thread and the Doppler velocity of the heated prominence plasma. A very good match is thus obtained with the observations. A sketch of the observed physical processes is given on the right. Adapted from \citet{okamoto2015} and \citet{antolin2015}.}
	\label{fig:patrick}
\end{figure}

Standing transverse waves can be considered as superposition of counterpropagating waves, thus these can generate a turbulent cascade as well. In closed loops, initial perturbations such as flares, low coronal eruptions or shock waves can easily excite standing kink oscillations. Due to the shear motions between the loops and external corona, the Kelvin-Helmholtz instability can thus be induced near the loop boundary \citep{2008ApJ...687L.115T,2014ApJ...787L..22A,2017ApJ...836..219A, 2015A&A...582A.117M}. Moreover, the generation of azimuthally polarised Alfv\'en waves due to resonant absorption increases the velocity shear with the external plasma, which can also enhance the instability \citep{antolin2019}. The effect of the instability is to generate transverse wave induced Kelvin-Helmholtz rolls (TWIKH rolls), extended along the magnetic field, as shown in Fig.~\ref{fig:KHI}. The wave energy can dissipate efficiently if the turbulent structures are sufficiently developed. Direct comparison of such a model with observations of a prominence \citep{okamoto2015,antolin2015}, gives an indication that the emission is indeed shifting from a temperature of $10^4$ K to $10^5$ K after the passing of the transverse wave, in accordance with the numerical model (see Fig.~\ref{fig:patrick}).
Recent studies by \citet{antolin2019} revealed that resonant absorption plays a key role in energizing and spreading the TWIKH rolls throughout a coronal loop.\par

The most basic setup that was used was a cylindrical loop with an initial velocity pulse, setting up a damped standing wave. This setup was used to show that the non-linearly developing KHI introduces an amplitude dependent damping \citep{magyar2016}, which matches reasonably well with observations \citep{2016A&A...585A.137G}. As intuitively understood, the magnetic twist provides a stabilising effect \citep{howson2017b, terradas2018}, which delays the development of the instability without completely suppressing it. This corresponds well with the predictions from analytical models \citep{hillier2019,barbulescu2019}. Moreover, the stabilising effect of viscosity and resistivity was investigated numerically \citep{howson2017,karampelas2019}. However, as explained in Sec.~\ref{sec:standing}, simple calculations using basic estimates show that the directly  observed standing waves in the solar corona do not seem to contain sufficient energy for heating it.

It was soon realised that the TWIKH rolls could play an important role if waves are driven at a loop footpoint. In that configuration, the TWIKH rolls provide an elegant mechanism to move the energy from the input scales, to the smaller length scales of the inertial range, and on to the dissipation scales at which the plasma is heated, just as Alfv\'en wave turbulence would do. This damping mechanism in the inertial range is independent of transport coefficients. In doing so, the energy cascade balances the input energy, creating a (statistically) steady state, indirectly feeding the driving energy into plasma heating. The steady state that is obtained in the numerical models could be a potential model for the observed decayless observations.

In the setup with the footpoint driver in a closed loop, \citet{karampelas2018} obtained a fully deformed loop cross-section. Although the turbulence significantly changes the initial equilibrium, the loop structure can still be observed in forward models, meaning that turbulent loops probability exist in the real solar corona. Further studies \citep{karampelas2019} considering gravity in a loop, stressed the importance of longitudinal stratification, which improves the efficiency of wave heating. The TWIKH roll formation dependence on the driver frequency was investigated by \citet{afanasyev2019}, who subsequently also considered the driving with a turbulent spectrum \citep{Afanasyev2020}.

\begin{figure}[t]
  \centering
  \medskip
  \includegraphics[width=0.9\textwidth]{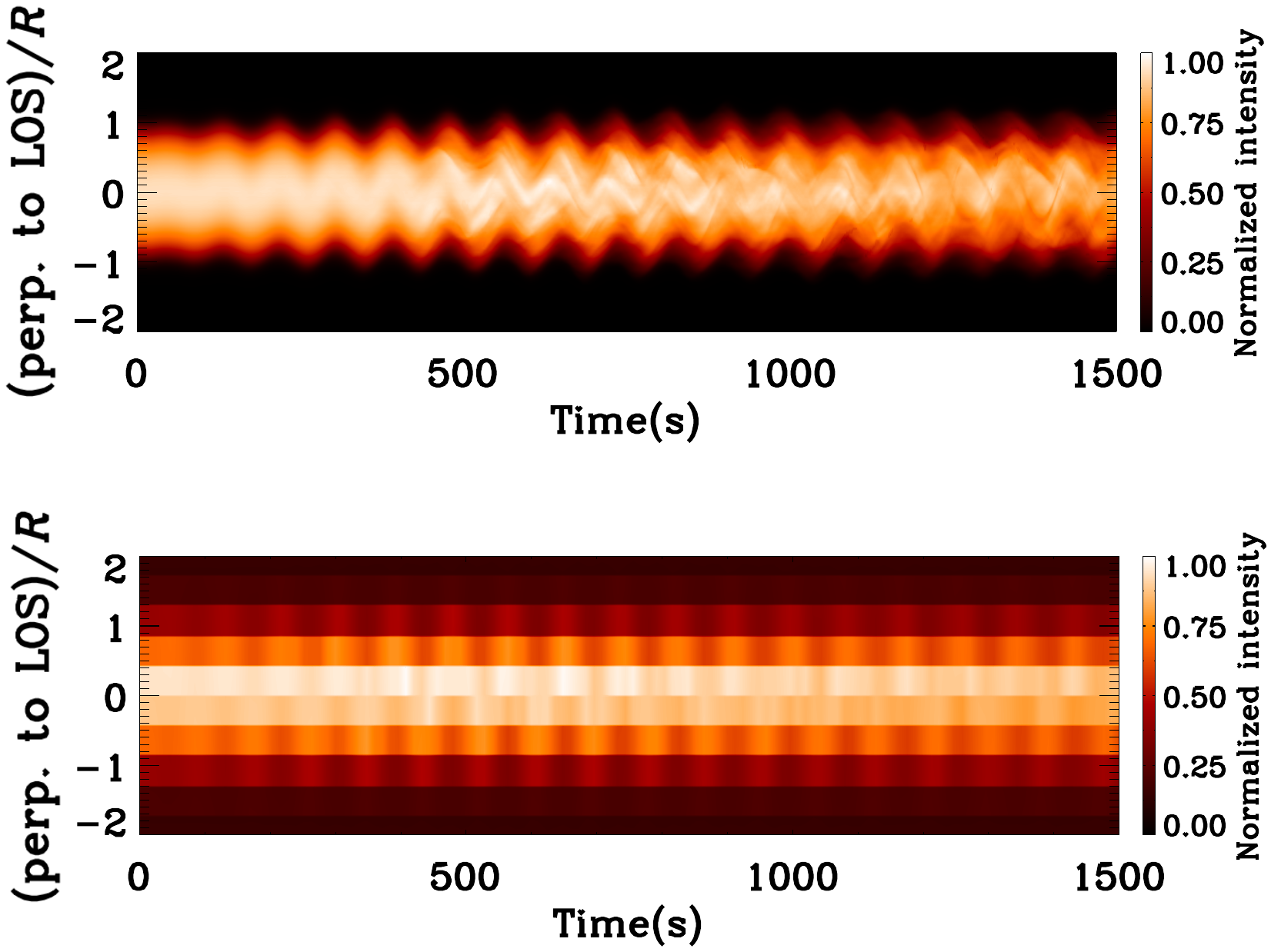}
  \caption{Forward-modelling results for the numerical models in the Fe  \uppercase\expandafter{\romannumeral 9} 171 $\AA$ line at the apex with an LOS angle of $45^\circ$. The upper row is obtained with the full numerical resolution and the lower row with a degraded resolution comparable to SDO/AIA. Figure taken from \citet{guo2019}.} 
  \label{fig_fomo}
\end{figure}

Considering the turbulent motions at the footpoint of a coronal loop anchored in the photosphere, \citet{guo2019} employed a mixed transverse and torsional driver at the loop footpoint, going beyond the implicitly excited Alfv\'en waves in the simulations of \citet{pascoe2011}. Comparing with a pure kink driven loop and a pure torsional driven loop, they found that the KHI eddies are fully developed in the whole loop cross-section and that internal energy and temperature in the mixed driven loop are higher. This means that the mixed modes lead to a more efficient dissipation in the turbulent state of plasma and that the KHI acts as an agent to dissipate energy in both wave modes. In addition, forward models showed very similar images to the observations of decayless oscillations. This means that the ubiquitous decayless oscillations could play an important role in coronal heating. However, as is shown in Figure \ref{fig_fomo}, due to the limited resolution of instruments, neither Alfv\'en modes nor the fine structures are observable. This means that although difficult to capture, Alfv\'en modes probably can coexist with kink modes, leading to enhanced heating.

A further study by \citet{guo2019b} examined how such a mixed mode driver affects a tightly packed multi-stranded loop. By depositing almost identical input energy flux at the footpoint of a multi-stranded loop and a density equivalent monolithic loop, \citet{guo2019b} found that the multi-stranded loop is more efficient in starting the heating process due to the quickly established turbulent state in this model.\\
Again using a model with footpoint driven loops, \citet{karampelas2019b} computed the dependence of the wave amplitude as they would be observed from forward modelling, on the driver amplitude. They found that there is only a weak dependence between both. Thus, they showed that the amplitude of the decayless waves is actually an unrealiable measure of the wave energy that is being dissipated in the solar corona, as studied in more detail by \citet{hillier2020}. \citet{karampelas2019b} concluded that the spectral line broadening is a much more accurate quantity to give a measure of the wave energy that is being dissipated. This may influence assessments of energy content and flux of standing or propagating kink waves (as e.g. in Sec.~\ref{sec:standing}).

Even though these models seem promising, at the time of submission, there is no published paper in which the non-linearly evolving, footpoint-driven transverse waves can compensate the radiative or thermal conduction cooling of the coronal plasma in a self-consistent model, contrary to DC heating models of the solar corona \citep[e.g.][]{2015ApJ...811..106H}. But, it should be noted that the 3D MHD DC models usually employ unrealistically high resistivity and viscosity, which are the sources for the heating in them.

\section{Critical Assessment \& Conclusions}\label{wh_conclusions}
This review provides an overview of the progress made in (roughly) the last decade in observations and models of wave heating in the solar corona. A lot of progress has been made in wave-based heating mechanisms using a  combination of advanced high-resolution satellite and ground-based solar observations, and advances in computational MHD modeling. \\
From an observational point of view, we now have detailed observations of several wave modes that could play a role in heating the corona. Recent multi-wavelength studies of coronal loops have not found evidence of any significant heating following large-amplitude kink oscillations \citep{2020A&A...638A..89G,2020ApJ...898..126P}. However, decayless kink oscillations and quasi-periodic propagating fast waves could potentially account for coronal losses in certain regions. We now have several observational characterisations of the wave energy content in the solar corona, based on imaging, spectroscopy or spectropolarimetry. In the chromosphere, a very large energy flux is found in waves, and if even only a fraction of that is dissipated in the corona, it is heated efficiently. In the corona itself, wave observations find energy fluxes comparable to the required heating rate, but also one or two orders of magnitude below it, particularly in the case of active regions. This wide disparity in observational estimates of coronal wave energy  points to a variety of wave activity manifestations, and  probably indicates that the issue of coronal wave energy requires further study. The latter point is also underlined by the discrepancy between the energy flux in numerical models and their forward modelling \citep{demoortel2012,vd2014,karampelas2019b}. 

From the modelling point of view, the computational models have been lifted to a new level in recent years  thanks to advancement in computational power. Now, 3D MHD simulations are the norm. In addition to previously considered AC heating mechanisms by non-linear Alfv\'en waves, these 3D MHD simulations show that turbulence and development of small scales is important. Simulations of Alfv\'en wave turbulence in (perpendicularly) uniform plasmas are able to maintain coronal temperatures, even in 3D configurations. In order to take into account the perpendicular density structuring, studies have focused on phase mixing and the transverse wave induced Kelvin-Helmholtz instability. Despite the strong effort in these last two subjects, still no self-consistent simulation exists in which a density enhanced coronal loop is maintained against cooling by radiation or conduction,  with resolved wave energy flux from the photosphere.

Self-consistent wave modelling studies are extremely demanding in computational power. The photospheric source (or sub-photopheric), the chromosphere (rich in wave processes due to its high- to low-$\beta$ variable conditions), the transition region (in which the Alfv\'en speed steepens leading to reflection, refraction and mode conversion) and the corona (in which small density inhomogeneities and wave-to-wave interaction are important) need to be included in the model. In addition, the resolution needs to be high enough to allow wave-associated instabilities and the turbulent spectrum to set in with the inertial range established for at least a decade in wave number. Through the decades, these limitations have favoured the investigation of wave dissipation in waveguides that are already well-defined (such as coronal loops) without the connection to other atmospheric layers \citep[the latter is only recently changing, e.g.][]{vandamme2020}. Although limiting, this approach has lead to an essential categorisation of wave processes and wave dissipation locally, and to phase relations between observables for a direct comparison with high resolution observations \citep[][chapter 6.1]{Hinode_10.1093/pasj/psz084}. It remains to be seen how these models connect to more advanced models of wave propagation in the lower atmosphere \citep[e.g.][]{2011ApJ...735...65F}.

This approach has also allowed to investigate the relative importance of specific wave mechanisms and the available energy at the coronal base \citep[e.g.][]{2019ApJ...871....3S}. Improvements in numerical techniques and increased computation power are coming close to allowing self-consistent modelling which incorporates all the relevant wave processes in a large scale model (such as a whole active region). This would allow to fully account for nonlinear processes and, in particular, global observables of wave heating models for an easier comparison with observations.

These recent developments have made it clear that the omnipresent coronal waves also strongly influence DC heating models. \\
Firstly, due to the observed continuous influx of Alfv\'enic waves into the corona, it is expected that the resulting heating is more continuous and less impulsive than that obtained from magnetic reconnection alone.  This long-standing assumption is only true if waves and reconnection act in a sufficiently independent way from each other. However, we can think of scenarios in which one facilitates the other. For instance, reconnection-based nanoflares in the corona are based on the braiding scenario, through which the magnetic field is stressed to a certain degree before releasing these stresses through component (small-angle) reconnection, thus leading to impulsive, middle to low-frequency events \citep[see chapter 6.2 in][]{Hinode_10.1093/pasj/psz084}. Moreover, transverse wave induced Kelvin-Helmholtz rolls (see Sec.~\ref{sec:kinkkhi}) create local twists and small-angle misalignments between the strands and, as in the magnetopause \citep{Nykyri_Otto_2004AnGeo..22..935N}, magnetic reconnection is expected. In this scenario, the heating from reconnection is expected to be impulsive and act only when the KHI is triggered. Although twists delay the formation of the KHI, they do not inhibit the instability (see Sec.~\ref{sec:kinkkhi}). Further investigation of this scenario is therefore needed.

Secondly, DC heating models are often based on stranded loop models, in which each magnetic field line is modelled independently from each other \citep{reale2014}. The turbulence appearing in Alfv\'en wave and kink wave models of loops are very efficient at mixing plasma perpendicular to the magnetic field \citep{magyar2016}. This mixing induces adiabatic expansion and contraction of plasma elements, leading to (apparent) cooling and heating on particular field lines, which is not taken into account if they are completely decoupled. Moreover, additional magnetic pressure enhancements are introduced, leading to a restructuring of the plasma in the single fieldline models. 

Thirdly, the nonlinearly induced perturbations due to Alfv\'en waves are not restricted to funnels and flux tubes with transverse or parallel structuring, but also feature in magnetic reconnection sites when Alfv\'en waves interact with a magnetic null point. As the Alfv\'en wave nears the magnetic null-point it induces magnetoacoustic waves which in turn contribute towards heating at the null point \citep{2003JGRA..108.1042G,2013A&A...555A..86T,2018MNRAS.479.4991S}. The fast and slow magnetoacoustic waves experience shocks in the vicinity of the null-point leading to enhanced dissipation there \citep{2009A&A...493..227M,2018MNRAS.479.4991S}. Thus, despite the Alfv\'en wave being unable to reach the null point while travelling along the separatrices, their coupling to the magnetoacoustic waves provides an agent to their gradual dissipation and contribution towards heating near X-points \citep{2004A&A...420.1129M,2006A&A...452..603M}.

Overall, we conclude this review with the following one-liner: Motivated by plentiful recent observations of coronal waves, AC heating models have taken a leap forward in recent years, but have not yet managed to self-consistently maintain loops at the observed million degree temperatures. These models make it worthwile to have a renewed look at DC heating models while including the cross-field effects induced by wave dynamics.

\section*{Acknowledgements}
This paper originated in discussions at ISSI-BJ.
T.V.D. was supported by the  European Research Council (ERC) under the European Union's Horizon 2020 research and innovation programme (grant agreement No 724326) and the C1 grant TRACEspace of Internal Funds KU Leuven (number C14/19/089).
H.T. is supported by NSFC Grants No. 11825301 and No. 11790304(11790300). 
P.A. acknowledges funding from his STFC Ernest Rutherford Fellowship (No. ST/R004285/2). Numerical computations were carried out on Cray XC50 at the Center for Computational Astrophysics, NAOJ.
D.J.P. was supported by the European Research Council (ERC) under the European Union's Horizon 2020 research and innovation programme (grant agreement No 724326).
L.O. acknowledges support by NASA grants  NNX16AF78G, 80NSSC18K1131 and NASA Cooperative Agreement NNG11PL10A to CUA.
I.A. was supported by project PGC2018-102108-B-I00 from Ministerio de Ciencia, Innovacion y Universidades and FEDER funds.
I.D.M. acknowledges support from the UK Science and Technology Facilities Council (Consolidated Grant ST/K000950/1), the European Union Horizon 2020 research and innovation programme (grant agreement No. 647214) and the Research Council of Norway through its Centres of Excellence scheme, project number 262622.
D.Y.K. acknowledges support from the STFC consolidated grant ST/T000252/1 and the budgetary funding of Basic Research program No. II.16.

%\section*{References}
\bibliographystyle{spbasic}
\bibliography{refs_arxiv,refs_review,kolotkov_heating_slow,aks_energy,Biblio,patbib,ofman_input,refs_djp}

\end{document}